\documentclass[twocolumn]{aastex631}

\DeclareRobustCommand{\VAN}[3]{#2}

\usepackage{CJK}
\usepackage{placeins}

\newcommand{\figdata}[2]{\begin{figure*}
    \centering
    \includegraphics[width=\textwidth]{tic#1.png}
    \caption{TESS data and amplitude spectra of #2 (TIC #1). The smoothed and fitted spectrum are shown in dashed and solid red lines.}
    \label{fig:#1}
\end{figure*}}

\usepackage{CJK}

\shorttitle{BSG Variability with TESS}
\shortauthors{Ma et al.}

\graphicspath{{./}{./plots/}{./plots/all_data/}}

\begin{document}
\begin{CJK*}{UTF8}{gbsn}

\title{Variability of Blue Supergiants in the LMC with TESS}
\correspondingauthor{Linhao Ma}
\email{lma3@caltech.edu}

\author[0000-0001-6117-5750]{Linhao Ma（马林昊）}
\affiliation{TAPIR, Mailcode 350-17, California Institute of Technology, Pasadena, CA 91125, USA}
\affiliation{Max-Planck-Institut f\"ur Astrophysik, Karl-Schwarzschild-Stra{\ss}e 1, 85741 Garching  bei M\"unchen, Germany}

\author[0000-0002-3054-4135]{Cole Johnston}
\affiliation{Max-Planck-Institut f\"ur Astrophysik, Karl-Schwarzschild-Stra{\ss}e 1, 85741 Garching  bei M\"unchen, Germany}
\affiliation{Department of Astrophysics, IMAPP, Radboud University Nijmegen, PO Box 9010, 6500 GL Nijmegen, The Netherlands}
\affiliation{Institute of Astronomy, KU Leuven, Celestijnenlaan 200D, 3001 Leuven, Belgium}

\author[0000-0003-4456-4863]{Earl Patrick Bellinger}
\affiliation{Max-Planck-Institut f\"ur Astrophysik, Karl-Schwarzschild-Stra{\ss}e 1, 85741 Garching  bei M\"unchen, Germany}
\affiliation{Department of Astronomy, Yale University, PO Box 208101, New Haven, CT 06520-8101, USA}
\affiliation{Stellar Astrophysics Centre, Aarhus University, 8000 Aarhus, Denmark}

\author[0000-0001-9336-2825]{Selma E. de Mink}
\affiliation{Max-Planck-Institut f\"ur Astrophysik, Karl-Schwarzschild-Stra{\ss}e 1, 85741 Garching  bei M\"unchen, Germany}
\affiliation{Anton Pannekoek Institute for Astronomy and GRAPPA, University of Amsterdam, NL-1090 GE Amsterdam, The Netherlands}

\begin{abstract}

The blue supergiant (BSG) problem, namely, the overabundance of BSGs inconsistent with classical stellar evolution theory, remains an open question in stellar astrophysics. Several theoretical explanations have been proposed, which may be tested by their predictions for the characteristic time variability. In this work, we analyze the light curves of a sample of 20 BSGs obtained from the Transiting Exoplanet Survey Satellite (TESS) mission. We report a characteristic signal in the low-frequency ($f\lesssim2\;\mathrm{day}^{-1}$) range for all our targets. The amplitude spectrum has a peak frequency at $\sim0.2\;\mathrm{day}^{-1}$, and we are able to fit it by a modified Lorentzian profile. The signal itself shows strong stochasticity across different TESS sectors, suggesting its driving mechanism happens on short ($\lesssim\mathrm{months}$) timescales. Our signals resemble those obtained for a limited sample of hotter OB stars and yellow supergiants, suggesting their possible common origins. We discuss three possible physical explanations: stellar winds launched by rotation, convection motions that reach the stellar surface, and waves from the deep stellar interior. The peak frequency of the signal favors processes related to the convective zone caused by the iron opacity peak, and the shape of the spectra might be explained by the propagation of high-order, damped gravity waves excited from that zone. We discuss the uncertainties and limitations of all these scenarios.

\end{abstract}


\section{Introduction}

\begin{figure}
    \centering
    \includegraphics[width=0.95\columnwidth]{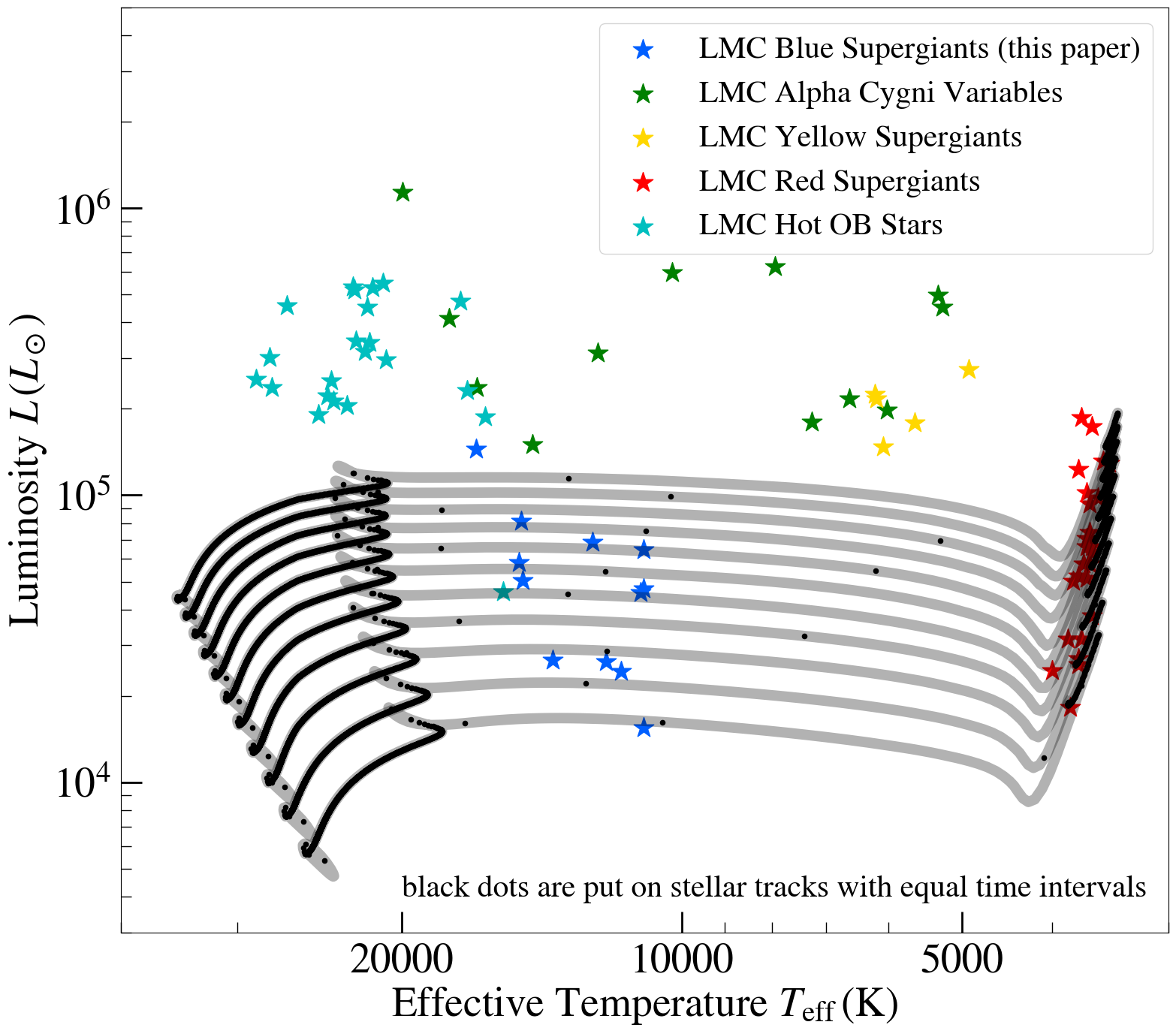}
    \caption{Blue supergiants (blue stars) form a distinct class in the Hertzsprung-Russell diagram, different from Alpha Cygni variables (green stars), yellow supergiants (yellow stars), red supergiants (red stars) or hot OB stars (cyan stars). When we put black dots with equal time intervals on the stellar tracks obtained from single stellar evolution model (from \protect\citealt{Bellinger2023}), BSGs lie in the region where the number of dots are orders of magnitude less than the dots on the main sequence, suggesting their occurrence rates should be orders of magnitude lower, inconsistent with the observed population. References: BSGs: 12 of the 20 stars in this paper with bolometric luminosity estimate; Alpha Cygnis: \protect\cite{vanLeeuwen1998}; YSGs: \protect\cite{Dorn-Wallenstein2020}; RSGs: \protect\cite{Neugent2020}; hot OB Stars: from \protect\cite{Bowman2019} with bolometric luminosity estimate.}
    \label{fig:HR}
\end{figure}

The formation pathways of blue supergiants (BSGs) are still not well understood. Despite being a unique class in the Hertzsprung--Russell (H--R) diagram, their observed population disagrees with predictions from the classical theory of massive star evolution (Figure~\ref{fig:HR}). Namely, if BSGs are massive single stars that have triggered hydrogen shell burning after central hydrogen depletion, their lifetime should be characterized by a thermal timescale on which their stellar envelope expands, typically $100-1000$ times faster than their main-sequence lifetime \citep{Hoyle1960,Hofmeister1964}. This means the occurrence rate of BSGs should be $100-1000$ times lower than massive main-sequence stars, clearly not consistent with their observed population \citep{Castro2014,Castro2018,deBurgos2023}. This discrepancy is known as the ``blue supergiant problem,'' and it remains one of the biggest open questions in the theory of stellar evolution \citep{Bellinger2023}.

Several alternative models for single stellar evolution have been proposed to solve the blue supergiant problem. Some authors argue that BSGs are actually central-helium-burning stars with smaller cores, and typically longer lifetimes. However, such theories usually require little or no mixing near the convective core boundary \citep{Walborn1987,Weiss1989,Schootemeijer2019}, a scenario disfavored by our current-day knowledge of massive stars \citep{Kaiser2020,Martinet2021}. Some other authors argue that BSGs could be core-hydrogen-burning stars that extend their main sequences to cooler temperatures and larger luminosities, where most BSGs lie on the H--R diagram \citep{Kaiser2020}. Such solutions typically require an excess amount of mixing beyond the convective core which produces large cores. They are the most satisfactory for hotter BSGs \citep{Brott2011}, but they appear to fail to match the very gradual observed decrease in the number of BSGs toward cooler temperatures \citep[e.g.\ ][]{Bellinger2023}.

In addition to single stellar evolution pathways, it has been proposed that binary interactions may produce long-lived BSGs, motivated by the fact that a large fraction of young massive stars are found in binaries where close interactions or mergers may occur \citep{Podsiadlowski1992,Kobulnicky2007,Sana2012,deMink2014}. Interactions between two unevolved stars may produce rapidly rotating stars \citep{DeMink2013}, which may experience extra mixing, resulting in an extended main sequence. If one of the stars has already evolved beyond the main sequence, it may give rise to a star with a smaller helium core compared to single stellar evolution products, either after accretion of hydrogen from a nearby star \citep{Braun1995}, or by merging  with a hydrogen-rich companion \citep{Vanbeveren2013,Justham2014}. Such a star could then evolve to a blue supergiant in the core-helium-burning phase \citep{Farrell2019}. Finally, stars that are partially stripped may also spend some time in the BSG \citep{Klencki2020,Laplace2020}, although this effect seems to work best at low metallicity. 

While it is uncertain which of these scenarios contribute to producing BSGs, they all concern different evolution pathways, which result in different internal stellar structures and core properties compared to expectations from classical stellar evolution theory. Thus, asteroseismology, which studies stellar pulsations whose frequencies are finely tuned by the internal structure of stars \citep{Aerts2010}, is a promising tool to investigate the structure, and hence, the origin of BSGs. In recent decades, there has been much interest in using asteroseismology to probe the structure and origin of BSGs from single-star evolutionary channels \citep{Stothers1968,Saio2006,Saio2011,Bowman2020review}. In particular, \citet{Saio2006} and \citet{Daszynska2013} identified an instability strip of gravity (g) mode pulsations in BSGs for shell-hydrogen and core-helium-burning models that arise due to trapping by the convective hydrogen burning shell and near core chemical gradient, respectively. Additional work by \citet{Godart2009} and \citet{Ostrowski2015} demonstrated that varying the physics, such as mass loss or including convective boundary mixing during the evolution of the main-sequence, can impact the presence of the chemical gradients and intermediate convective zones required to trap and excite modes in BSG models. \citet{Saio2013} and \citet{Ostrowski2015} further discussed the excitation of pulsations in models before and after core helium ignition, indicating that they produce different pulsation periods and excited mode spectra. While \citet{Daszynska2013} argue that regular oscillation patterns are not excited in BSGs, \citet{Bowman2019} demonstrate that, if observed, g mode period spacing patterns can be used to discriminate between BSGs crossing the Hertzsprung-gap and those experiencing blue loops. Recently, \cite{Bellinger2023} made the first predictions of the asteroseismic properties of BSGs that result from binary evolutionary channels. They further demonstrate that asteroseismic signals may help to distinguish different single star an binary evolution BSG formation scenarios. Crucially, several of these studies predict the excitation of a dense spectrum of g mode pulsations with frequencies between 0.1 and 1 day$^{-1}$ \citep{Saio2006,Daszynska2013,Ostrowski2015}.

In the past decades, several space missions have delivered high-precision time series measurements for tens of thousands of stars, including CoRoT \citep{Auvergne2009}, {\em Kepler}/K2 \citep{Borucki2010} and the Transiting Exoplanet Survey Satellite (TESS; \citealt{TESS}). While a handful of samples of BSG variability has been discovered with CoRoT and {\em Kepler}/K2 photometry (see, e.g., \citealt{Aerts2010,Aerts2017,Aerts2018,Sanchez2023}), the all-sky TESS mission offers a unique opportunity for a systematic study for BSGs due to its large field of view that covers hundreds of potential candidates. This has motivated recent work to investigate the variability of BSGs with TESS data. Specifically, \citet{Bowman2019,Bowman2019b,Bowman2020} have identified a stochastic low-frequency (SLF) excess in both main-sequence OB stars and BSGs. This SLF feature is characterized by an increase in the overall amplitude in the periodogram toward lower frequencies, with a constant amplitude profile as the frequency tends toward zero. \citet{Bowman2019} argue that this feature is the signature of internal gravity waves (IGWs) generated by core convection \citep{Rogers2013,Shiode2013}. However, this claim is contested. Specifically, \citet{Cantiello2021} and \citet{Schultz2023a,Schultz2023b} demonstrate that these signals can be caused by subsurface convective zones and rotationally modulated stellar winds launched at the surface. Furthermore, \citet{Anders2023} use 3D hydrodynamical simulations to argue that IGWs generated by core convection should not have observable amplitudes at the stellar surface. Despite the increased attention in recent years, it is clear that more observations are required to identify the causes of the various signals seen in BSGs and then interpret them.

In this work, we analyze the light curves of a sample of 20 BSGs in the Large Magellanic Cloud (LMC). We find a characteristic time variability signal in our sample that universally appears in LMC BSGs. The signal shows similarity to the low-frequency variability of other massive pulsators, suggesting they may have similar origins. The manuscript is organized as follows: We describe our sample selection procedure in Section \ref{sec:method} and present our results in Section \ref{sec:results}. We compare our signals to the low-frequency variability of other systems in Section \ref{sec:compare}, and in Section \ref{sec:discussions} we discuss the possible physical origins of our signal. We conclude in Section \ref{sec:conclusions}.

\begin{figure}
    \centering
    \includegraphics[width=\columnwidth]{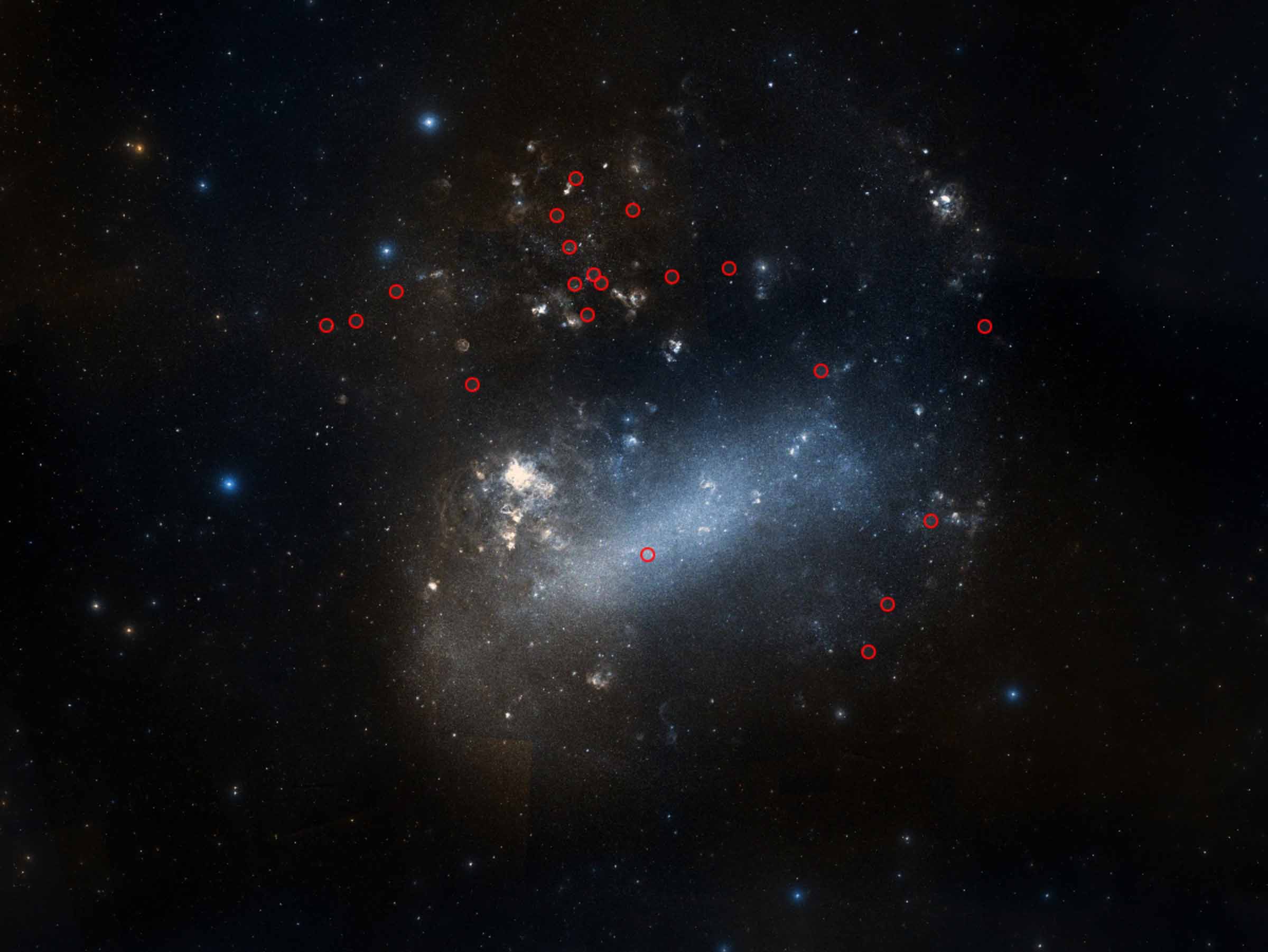}
    \caption{Locations of our BSG sample (red circles) in the Large Magellanic Cloud (obtained with the colored Digitized Sky Survey). The figure is made with SIMBAD \protect\citep{SIMBAD}.}
    \label{fig:stars}
\end{figure}

\section{Sample Selection}
\label{sec:method}

\begin{table*}
    \centering
    \fontsize{7.3}{10.5}\selectfont
    \begin{tabular}{llllll}
        \hline
         Object Name & TIC Number & $T_\mathrm{eff}\;(\mathrm{K})$ & $\log g\;(\mathrm{dex})$ & $v\sin i\;(\mathrm{km/s})$ & TESS Observing Sectors \\
         \hline
         SK -68 53 &  31179797 & $13770\pm740$ & $2.09\pm0.12$ & $47.5\pm11.5$ & 20 sectors: 27-31, 33, 34, 36-39, 61-69 \\
         SK -66 125 & 425086354 & $14830\pm760$ & $2.34\pm0.13$ & $36.5\pm11.5$ & 18 sectors: 27-35, 37-39, 62-64, 66-68 \\
         SK -67 283 & 31511729 & $14880\pm800$ & $2.26\pm0.14$ & $46.5\pm11.9$ & 12 sectors: 28, 29, 31, 32, 35, 38, 39, 61, 66-69 \\
         SK -69 31 &  30190076 & $11830\pm600$ & $1.7\pm0.12$ & $44.0\pm9.7$ & 20 sectors: 27-39, 61, 63-67, 69 \\
         SK -67 275 & 389864558 & $14960\pm890$ & $2.23\pm0.16$ & $63.5\pm12.1$ & 10 sectors: 27, 29, 30, 33, 36, 37, 39, 65, 68, 69 \\
         SK -66 142 & 276860494 & $13640\pm740$ & $2.0\pm0.12$ & $49.0\pm11.4$ & 19 sectors: 27, 28, 30-35, 37, 38, 61-69 \\
         HD 269639 &  287400996 & $11000\pm670$ & $1.7\pm0.15$ & $26.5\pm15.0$ & 19 sectors: 27, 28, 30, 31, 33-39, 61, 62, 64-69 \\
         SK -67 7 &   29987961 & $12460\pm760$ & $1.83\pm0.14$ & $46.5\pm11.8$ & 19 sectors: 27, 28, 30, 31, 33-39, 61-64, 66-69 \\
         SK -67 279 & 31311824 & $11330\pm600$ & $1.9\pm0.15$ & $23.0\pm10.2$ & 12 sectors: 29, 30, 32, 33, 35, 36, 38, 39, 61, 67-69 \\
         SK -70 31 &  30534618 & $12060\pm790$ & $1.9\pm0.15$ & $33.0\pm10.6$ & 21 sectors: 27-39, 61-63, 65-69 \\
         SK -68 152 & 389366376 & $11000\pm530$ & $2.01\pm0.13$ & $34.0\pm8.0$ & 16 sectors: 27, 29, 30, 32, 33, 35-39, 61, 64-66, 68, 69 \\
         SK -67 88 &  179638852 & $11220\pm600$ & $1.78\pm0.13$ & $44.0\pm10.0$ & 19 sectors: 27, 28, 30, 31, 33-39, 61, 62, 64-69 \\
         HD 269721 &  425084965 & $11080\pm600$ & $1.7\pm0.13$ & $40.0\pm8.9$ & 19 sectors: 27, 28, 30-38, 61, 62, 64-69 \\
         HD 269510 &  373682056 & $11000\pm620$ & $1.7\pm0.14$ & $39.0\pm10.6$ & 20 sectors: 27-30, 32, 33, 35-39, 61-69 \\
         SK -66 92 &  373845622 & $13480\pm750$ & $1.92\pm0.11$ & $51.5\pm11.0$ & 17 sectors: 27-35, 37, 38, 61, 62, 64, 66-68 \\
         SK -67 151 & 391809264 & $11000\pm510$ & $2.24\pm0.15$ & $8.0\pm19.8$ & 18 sectors: 27-31, 33, 34, 36-39, 61-63, 65-67, 69 \\
         SK -70 26 &  30403638 & $11630\pm570$ & $1.9\pm0.13$ & $39.5\pm9.8$ & 20 sectors: 27, 29-37, 39, 61-69 \\
         SK -67 171 & 425084139 & $12650\pm740$ & $1.8\pm0.12$ & $47.5\pm11.5$ & 16 sectors: 27, 28, 30, 31, 33, 34, 36-39, 61, 62, 65-67, 69 \\
         SK -67 133 & 287401176 & $16630\pm850$ & $2.28\pm0.12$ & $53.5\pm10.4$ & 29 sectors: 1, 3-11, 13, 27, 28, 30, 31, 33-39, 61, 62, 64-67, 69 \\
         SK -67 72 &  179038240 & $12480\pm750$ & $1.84\pm0.14$ & $46.5\pm11.8$ & 19 sectors: 27, 28, 30-38, 61-64, 66-69 \\
\hline
    \end{tabular}
    \caption{Names, Tess Input Catalog (TIC) numbers, spectroscopic data and TESS observing sectors of our BSG sample. $T_\mathrm{eff}$, $\log g$ and $v\sin i$ are taken from spectroscopic measurements from \protect\cite{Serebriakova2023} (UVES preferred when measurements from both instruments available), where we quoted their $v\sin i_\mathrm{SP}$ for references.}
    \label{tab:samples}
\end{table*}

We selected our sample from the 124 OB pulsators with the spectroscopic measurements in \cite{Serebriakova2023}, who obtained their data with the Ultraviolet and Visual \'Echelle Spectrograph ({\small UVES}; \citealt{Dekker2000}) and the Fiber-fed Extended Range Optical Spectrograph ({\small FEROS}; \citealt{Kaufer1999}) instruments on board the UT2@VLT and MPG/ESO 2.2\,m telescopes. Their sample is largely based on the massive stars in \cite{Bowman2019} and \cite{Pedersen2019} with photometric variability detected in the first sector(s) of TESS data, and they carried out high-resolution ($R\gtrsim40000$), high-signal-to-noise-ratio ($S/N\sim100$) two-epoch spectroscopy such that binaries were excluded. The targets all locate in the Southern Continuous Viewing Zone (CVZ-S) of TESS and its vicinity, with high-duty-cycle photometric data of between approximately 200 days and 1 year in duration.

Out of the 148 pulsators, we selected 41 BSG candidates on the spectroscopic H--R diagram, with $3.2<\log{(\mathcal{L}/\mathcal{L}_\odot})<4.0$ and $T_\mathrm{eff}<30000\,\mathrm{K}$, where $\mathcal{L}:= T_\mathrm{eff}^4/g$ is the spectroscopic luminosity \citep{Langer2014}. The more luminous stars may still be BSGs of higher masses, yet we exclude them from our candidates since their evolution stages are even less clear (see Section \ref{sec:compare} for a comparison between our sample and a limited sample of hotter B stars). Bolometric corrections turn out to be difficult for these candidates since many of them do not have reliable measurements of extinction, hence we did not select our sample from their real luminosities.

As each TESS pixel has an angular size of $21''$, stars around our targets could cause contamination to the light curve, and this is especially an issue when the star locates in crowded regions of the sky \citep{Pedersen2023}. To exclude false signals caused by nearby stars, we used the {\small LATTE} package \citep{LATTE} to obtain star fields for all our candidates from both TESS and {\em Gaia} Data Release 2 \citep{GaiaDR2}. We compared the star fields and only kept the candidates without any bright (defined by the difference of magnitudes $\Delta m<4$) neighbors in the same TESS pixel as them. After such treatment, 20 stars are left in our final sample. It turns out that they are all in the LMC (Figure~\ref{fig:stars}), and their spectroscopic properties are listed in Table \ref{tab:samples}.

\begin{figure*}
    \centering
    \includegraphics[width=\textwidth]{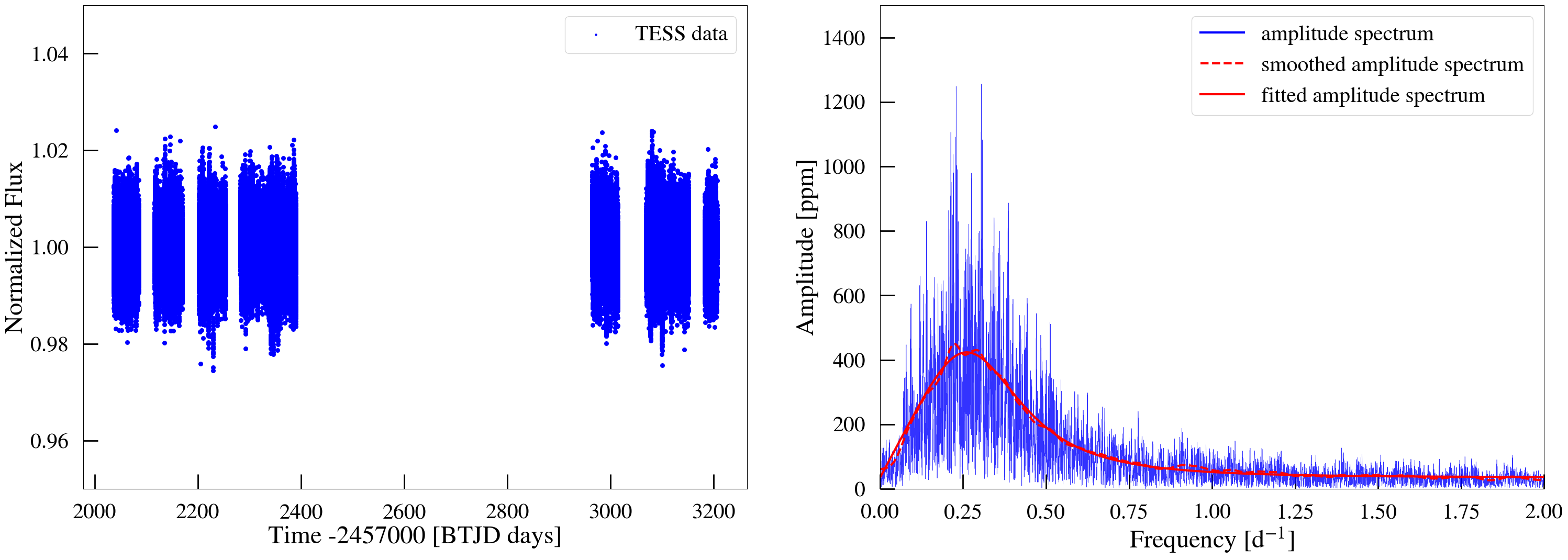}
    \caption{TESS data and amplitude spectra of SK -67 171 (TIC 425084139). The smoothed and fitted spectrum are shown in dashed and solid red lines. The amplitude spectrum shows a pattern with a peak frequency at $\sim 0.25\;\mathrm{day}^{-1}$, and can be fitted by a modified Lorentzian profile (see main text).}
    \label{fig:data}
\end{figure*}

We used the Python package Lightkurve \citep{Lightkurve} to download the TESS light curves for our sample from the Mikulski Archive for Space Telescopes (MAST\footnote{https://archive.stsci.edu/}), using the data processed by a pipeline developed by the Science Processing Operations Center (SPOC; \citealt{Jenkins2016}; for its validity in processing low-frequency signals, see discussions in Appendix \ref{app:spoc}). We used Astropy to analyze the data \citep{astropy:2013, astropy:2018, astropy:2022}. Our sample has 10-29 observing sectors available from TESS, with typical observation period of $\sim 3$ years (sector 27-69, from July 5, 2020 to September 30, 2023). Their details are summarized in Table \ref{tab:samples}. All the {\it TESS} data used in this paper can be found in MAST: \dataset[https://doi.org/10.17909/df38-ax53]{https://doi.org/10.17909/df38-ax53}.

\section{Results}

\label{sec:results}

Here we present our results of identifying and characterizing the variability for the BSGs in our study.

\subsection{Universal Signal in BSGs}

Figure~\ref{fig:data} shows the TESS data and periodogram of one typical example system in our sample: SK -67 171 (TIC 425084139). The results for all our sample can be seen from Figures \ref{fig:31179797}-\ref{fig:179038240} in Appendix \ref{app:all_data}. While we do not see any significant individual modes from the periodogram of this particular system, we do identify a low-frequency ($f<2\,\mathrm{d}^{-1}$) variability signal from the Fourier transform of the time series, as shown. The signal goes to zero amplitude toward high and low frequencies, with a characteristic peak frequency at $\sim0.25\,\mathrm{d}^{-1}$. We see in Appendix \ref{app:all_data} that this signal appears in all of the targets in our sample. We suggest that this signature may be a common, or potentially ubiquitous phenomenon in BSGs. We note that other works have previously identified a variety of both stochastic and periodic variability in hot, main-sequence and evolved stars \citep[e.g., ][]{vanLeeuwen1998,Saio2006,Bowman2019}. However, this is the first reporting of a signature that tends to zero amplitude toward higher and lower frequencies in a sample of BSGs.

The high-frequency tail above the peak frequency in our signal looks very similar to a characteristic signal that has already been widely discussed in hot massive stars, commonly referred as the stochastic low-frequency (SLF) variability or ``{\em red noise}''  \citep{Bowman2019,Bowman2019b,Bowman2020,Szewczuk2021,Bowman2022,Dorn-Wallenstein2022}. We note that, however, the signal pattern in our sample is distinct as it has decaying power toward zero frequency, while the red noise signal mostly concerns the high frequency tail. As the red noise signal is usually fitted with a super Lorentzian profile (e.g., \citealt{Kallinger2014}), we are motivated to characterize our amplitude spectra with the following modified Lorentzian profile:

\begin{equation}
\label{eq:fit}
P(\nu)=W_0+\alpha_0\frac{(\nu/\nu_\mathrm{char})}{1+(\nu/\nu_\mathrm{char})^\gamma}\,,
\end{equation}
which is a linear profile times a super-Lorentzian profile with three parameters ($\nu_\mathrm{char}$, $\alpha_0$ and $\gamma$), plus a background noise term $W_0$. This profile captures the shape of the signal with a peak frequency $\nu_\mathrm{peak}=(\gamma-1)^{-1/\gamma}\nu_\mathrm{char}$ and a tendency toward zero at both high and low frequencies. It also resembles the high frequency tail of super-Lorentzian profile  above the peak frequency.

We smoothed our amplitude spectra with a window size of $0.1\,\mathrm{d}^{-1}$ and fitted the smoothed spectra with the above function using a nonlinear least squares method with the \texttt{curve\_fit} function in the Python package ScipPy \citep{scipy}. The fitting parameters we found are shown in Table \ref{tab:fit_params} and an example fitting curve is shown in Figure~\ref{fig:data} (for all other stars in our sample, see Figures \ref{fig:31179797}-\ref{fig:179038240}). We found that our fitting profile matches the smoothed spectrum very well, with the background-noise term $W_0$ at least one order-of-magnitude lower than the characteristic amplitude term $\alpha_0$. The physical peak frequency $\nu_\mathrm{peak}$ is around $0.2$ days, and the power-law index $\gamma$ ranges from $2$ to $5$, revealing a universal pattern in all our stars with small dispersion. We note that, however, the profile is not physically motivated and we did not find any strong correlations between the fitting parameters and the spectroscopic properties of our sample, which may give insights into the physical origins of the signal. We confirmed that our fitting parameters are robust against tests with different choices of smoothing windows sizes.

\subsection{Time Variability of the Signal}
\label{sec:stochaticity}

\begin{table}
    \centering
    \begin{tabular}{lccccc}
    \hline
    Object & $W_0$ & $\alpha_0$& $\nu_\mathrm{char}\;$ & $\gamma$ & $\nu_\mathrm{peak} $ \\
      &  (ppm) &(ppm)& $(\mathrm{d}^{-1})$ &  & $ (\mathrm{d}^{-1})$ \\
    \hline
         SK -68 53 & 7.0 & 864.67 & 0.235 & 2.6 & 0.195 \\
SK -66 125 & 27.1 & 727.40 & 0.256 & 2.8 & 0.209 \\
SK -67 283 & 10.5 & 1108.24 & 0.184 & 2.3 & 0.164 \\
SK -69 31 & 47.7 & 1021.72 & 0.236 & 4.1 & 0.179 \\
SK -67 275 & 1.6 & 889.66 & 0.195 & 2.2 & 0.178 \\
SK -66 142 & 17.2 & 953.07 & 0.277 & 3.0 & 0.219 \\
HD 269639 & 36.2 & 672.25 & 0.224 & 3.8 & 0.171 \\
SK -67 7 & 9.0 & 208.45 & 0.344 & 2.7 & 0.284 \\
SK -67 279 & 20.9 & 524.19 & 0.315 & 3.0 & 0.251 \\
SK -70 31 & 14.3 & 472.39 & 0.321 & 3.2 & 0.251 \\
SK -68 152 & 22.5 & 358.37 & 0.298 & 5.1 & 0.226 \\
SK -67 88 & 35.6 & 749.95 & 0.292 & 4.3 & 0.221 \\
HD 269721 & 43.3 & 896.16 & 0.200 & 3.9 & 0.152 \\
HD 269510 & 16.0 & 331.78 & 0.117 & 2.5 & 0.099 \\
SK -66 92 & 30.0 & 1143.31 & 0.277 & 3.4 & 0.214 \\
SK -67 151 & 25.2 & 299.23 & 0.298 & 4.8 & 0.226 \\
SK -70 26 & 15.8 & 375.17 & 0.325 & 3.1 & 0.256 \\
SK -67 171 & 35.0 & 662.57 & 0.345 & 4.4 & 0.261 \\
SK -67 133 & 9.4 & 1004.97 & 0.271 & 2.7 & 0.223 \\
SK -67 72 & 21.8 & 721.37 & 0.301 & 3.4 & 0.233 \\
\hline
    \end{tabular}
    \caption{Fitting parameters and the physical peak frequency $\nu_\mathrm{peak}=(\gamma-1)^{-1/\gamma}\nu_\mathrm{char}$ for the modified Lorentzian profile (Equation \ref{eq:fit}) that best matches our signal.}
    \label{tab:fit_params}
\end{table}

The above analysis was carried out with the full TESS data obtained for our BSG sample, across multiple sectors. With an average observing time of 27 days per sector, the full TESS light curve gives us a frequency resolution of $0.001-0.005$ d$^{-1}$. We note that, however, the signal actually varies in time as well, as seen by analyzing each observed sector individually.

\begin{figure}
    \centering
   \includegraphics[width=\columnwidth]{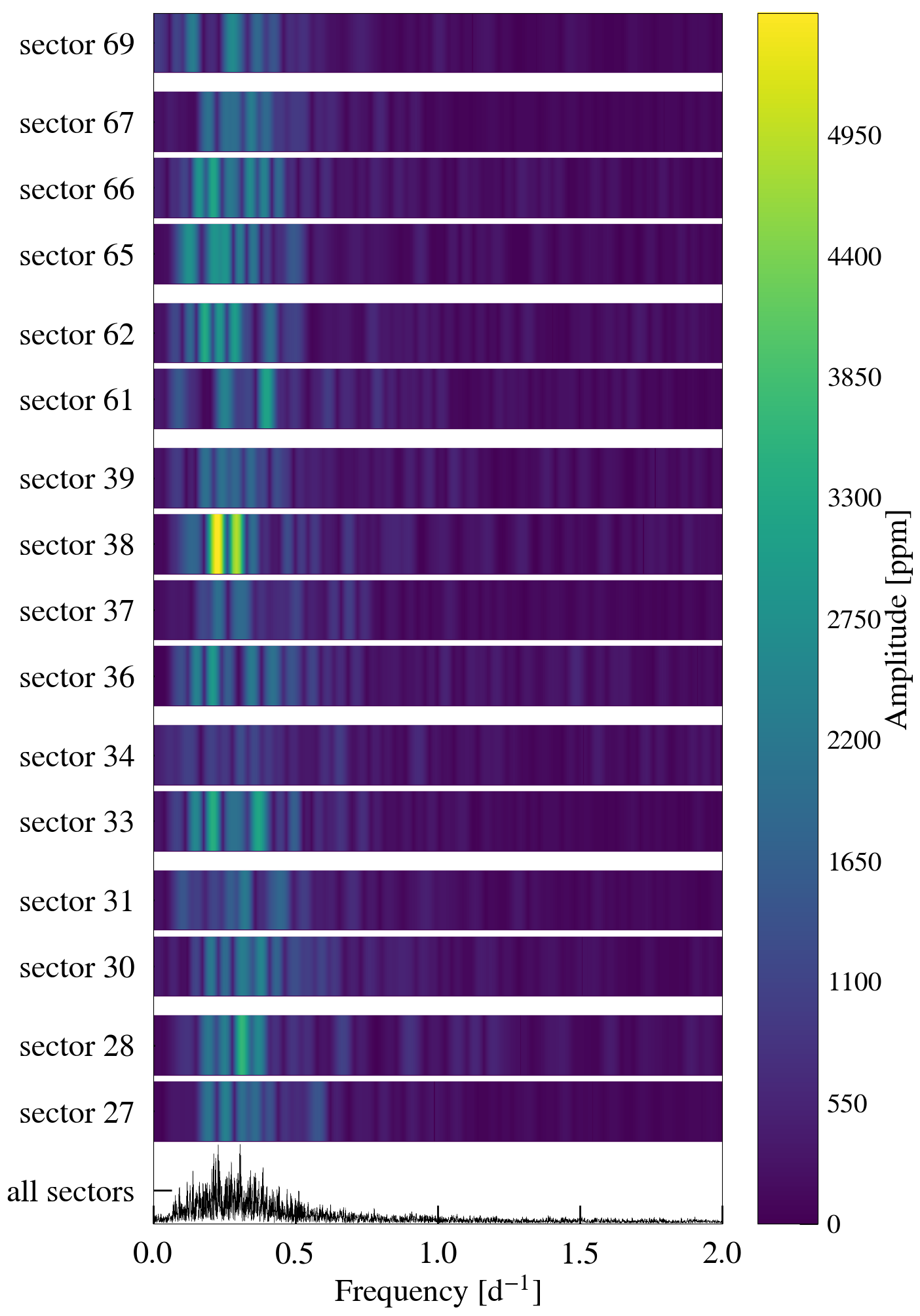}
    \caption{The amplitude spectrum from individual TESS sectors for BSG SK -67 171. Each bar represents an amplitude spectra from one sector, with the color mapping the amplitude at different frequencies (such that a ``lighter'' color region represents a peak at that frequency). The full amplitude spectra combining all sectors are shown in the bottom for reference. We see clear time variability of the spectra across different sectors.}
    \label{fig:sector_spectrum}
\end{figure}

Figure~\ref{fig:sector_spectrum} shows the amplitude spectrum obtained from individual sectors for the example BSG SK -67 171. Each bar represents an amplitude spectra from one sector, with the color mapping the amplitude at different frequencies (such that a lighter color region represents a peak at that frequency). The full amplitude spectra combining all sectors are shown at the bottom for reference. We see that the shape of the amplitude spectrum for individual sectors is different from each other and the full amplitude spectra. Some features in one sector are preserved in the next sector (e.g., in sectors 27 and 28), while some only return after a few sectors (e.g., in sectors 33 and 36), or never repeats. We note that as the frequency resolution of individual sectors is as low as $\Delta f\sim1/27\,\mathrm{d}=0.04\,\mathrm{d}^{-1}$, we cannot identify any of the features as distinct modes excited. Nevertheless, we found similar variability of the features for all our samples.

\begin{table*}
    \centering
    \begin{tabular}{lccccccc}
    \hline
    Object & $T_\mathrm{eff,spec}\;(\mathrm{K})$ & $\log(L_\mathrm{bol}/L_\odot)$ & $R_\mathrm{bol}/R_\odot$ & $\nu_\mathrm{peak}\;(\mathrm{d}^{-1})$ & $v_\mathrm{eq,derive}$ (km/s)& $v\sin i$ (km/s) & $(\sin i)_\mathrm{derive}$\\
    \hline
         SK -68 53 & 13770 & 4.425 & 28.7 & 0.195 & 283.4 & 47.5 & 0.17 \\
SK -66 125 & 14830 & 4.703 & 34.0 & 0.209 & 359.2 & 36.5 & 0.10 \\
SK -67 283 & 14880 & 4.909 & 42.8 & 0.164 & 354.9 & 46.5 & 0.13 \\
SK -67 275 & 14960 & 4.765 & 35.9 & 0.178 & 323.7 & 63.5 & 0.20 \\
HD 269639 & 11000 & 4.674 & 59.8 & 0.171 & 518.5 & 26.5 & 0.05 \\
SK -70 31 & 12060 & 4.421 & 37.2 & 0.251 & 472.9 & 33.0 & 0.07 \\
SK -68 152 & 11000 & 4.191 & 34.3 & 0.226 & 392.7 & 34.0 & 0.09 \\
HD 269721 & 11080 & 4.660 & 58.0 & 0.152 & 447.0 & 40.0 & 0.09 \\
HD 269510 & 11000 & 4.809 & 69.9 & 0.099 & 351.7 & 39.0 & 0.11 \\
SK -70 26 & 11630 & 4.387 & 38.5 & 0.256 & 498.0 & 39.5 & 0.08 \\
SK -67 133 & 16630 & 5.162 & 45.9 & 0.223 & 517.4 & 53.5 & 0.10 \\
SK -67 72 & 12480 & 4.836 & 56.0 & 0.233 & 660.1 & 46.5 & 0.07 \\
\hline
    \end{tabular}
    \caption{The derived bolometric luminosity, radius and equatorial velocity of 12 BSGs in our sample. The derived equatorial velocities are not compatible with spectroscopic $v\sin i$ measurements, typically one order of magnitude too large.}
    \label{tab:eq_v}
\end{table*}

As each TESS sector only observes for a period of $\sim 27\,\mathrm{days}$, such time variability in our signal suggests that the physical mechanism related to them must be very short, occurring on timescales shorter than $\sim$ 30 days. If the signal is stochastically excited from the stellar interior (see discussions in Section \ref{sec:modes}), the driving mechanisms must happen on such short timescales as well, and the characteristic signals they excite must be at least equally short-lived. This put additional constraints on understanding the physics behind them.

\section{Comparison to Low-Frequency Variability in Other LMC Stars}
\label{sec:compare}

Low-frequency photometric variability has been found and discussed in different kinds of massive stars on the H--R diagram, including hot OB stars \citep{Bowman2019b} and evolved yellow supergiants \citep{Dorn-Wallenstein2019,Dorn-Wallenstein2020}. To look for their similarities and possible common physical origins, we analyze a limited sample of other LMC stars and compare them with our BSG sample.

\begin{figure*}
    \centering
    \includegraphics[width=0.93\textwidth]{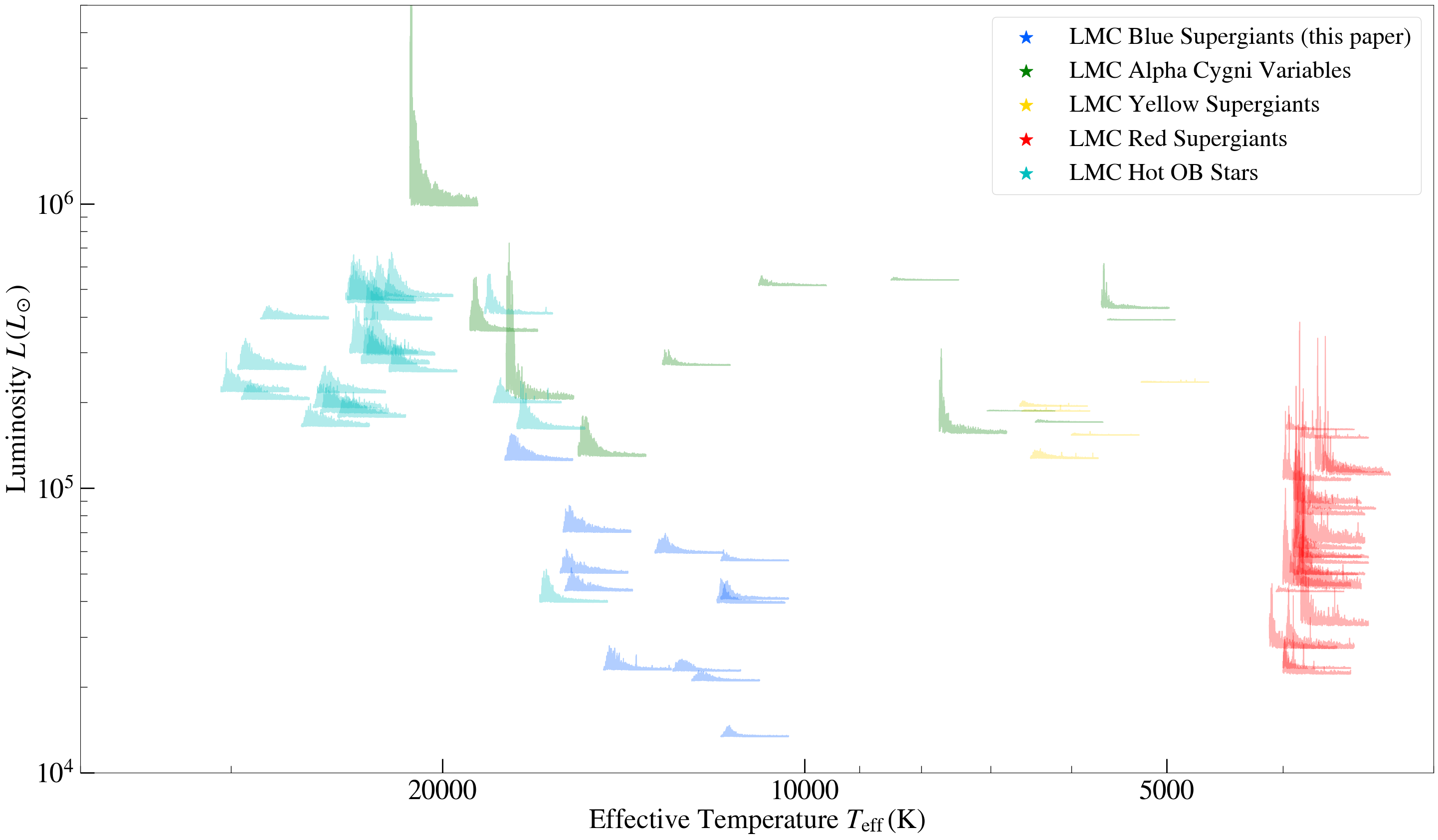}
    \caption{The amplitude spectra of 12 BSGs in the $0-2\,\mathrm{d}^{-1}$ frequency range plotted on their locations on the H--R diagram, compared to other types of stars (see Figure~\ref{fig:HR}). We see similar kind of variability in hotter OB stars and around half of the Alpha Cygni variables, suggesting they may share similar origins. References: BSGs: 12 of the 20 BSG in this paper with bolometric luminosity estimate; Alpha Cygnis: \protect\cite{vanLeeuwen1998}; YSGs: \protect\cite{Dorn-Wallenstein2020}; RSGs: \protect\cite{Neugent2020}; hot OB Stars: 22 stars from \protect\cite{Bowman2019} with bolometric luminosity estimate and spectroscopy measurements from \cite{Serebriakova2023}.}
    \label{fig:HR_periodogram}
\end{figure*}

In Figure~\ref{fig:HR_periodogram} we show the periodograms (between $0-2\,\mathrm{d}^{-1}$) for various types of stars with their positions on the H--R diagram. This includes 12 BSGs from this study with bolometric luminosity estimates (listed in Table \ref{tab:eq_v}), 5 yellow supergiants from \cite{Dorn-Wallenstein2020}, 23 red supergiants from \cite{Neugent2020}, 12 alpha Cygni variables from \cite{vanLeeuwen1998} and 22 hot OB stars from the TESS OB sample in \cite{Bowman2019b} with bolometric luminosity estimates and spectroscopy measurements from \cite{Serebriakova2023}. The data are acquired with TESS with similar analysis as described in \ref{sec:method}. With this limited sample, we clearly see similarities between the variability of hot OB stars and our BSG sample. Some of the alpha Cygni variables and yellow supergiants resemble the the signal we found in BSGs, yet with much lower power, while some other alpha Cygni variables show a massive peak at very low frequency with no clear turnover. The red supergiants, on the other hand, form the most distinct class on the H--R diagram and typically show some multi-peak pattern in the periodogram, very different from BSGs.

Previous photometric analyses on OB stars and yellow supergiants mostly interpret the low-frequency variability as ``red noise'', characterized by some function that levels off at zero frequency \citep{Bowman2019b,Dorn-Wallenstein2019,Dorn-Wallenstein2020}, which is clearly not what we see. This is probably because these authors mostly carried out their time series analysis for light curves shorter than $\lesssim 100\,\mathrm{days}$, such that the characteristic low-frequency turnover cannot be well resolved in Fourier space. Nevertheless, the existing theoretical explanations for their signals are still relevant to our sample, given the similarities between these signals, which is what we probe in the next section.

\section{Discussions}
\label{sec:discussions}

\begin{figure}
    \centering
    \includegraphics[width=0.95\columnwidth]{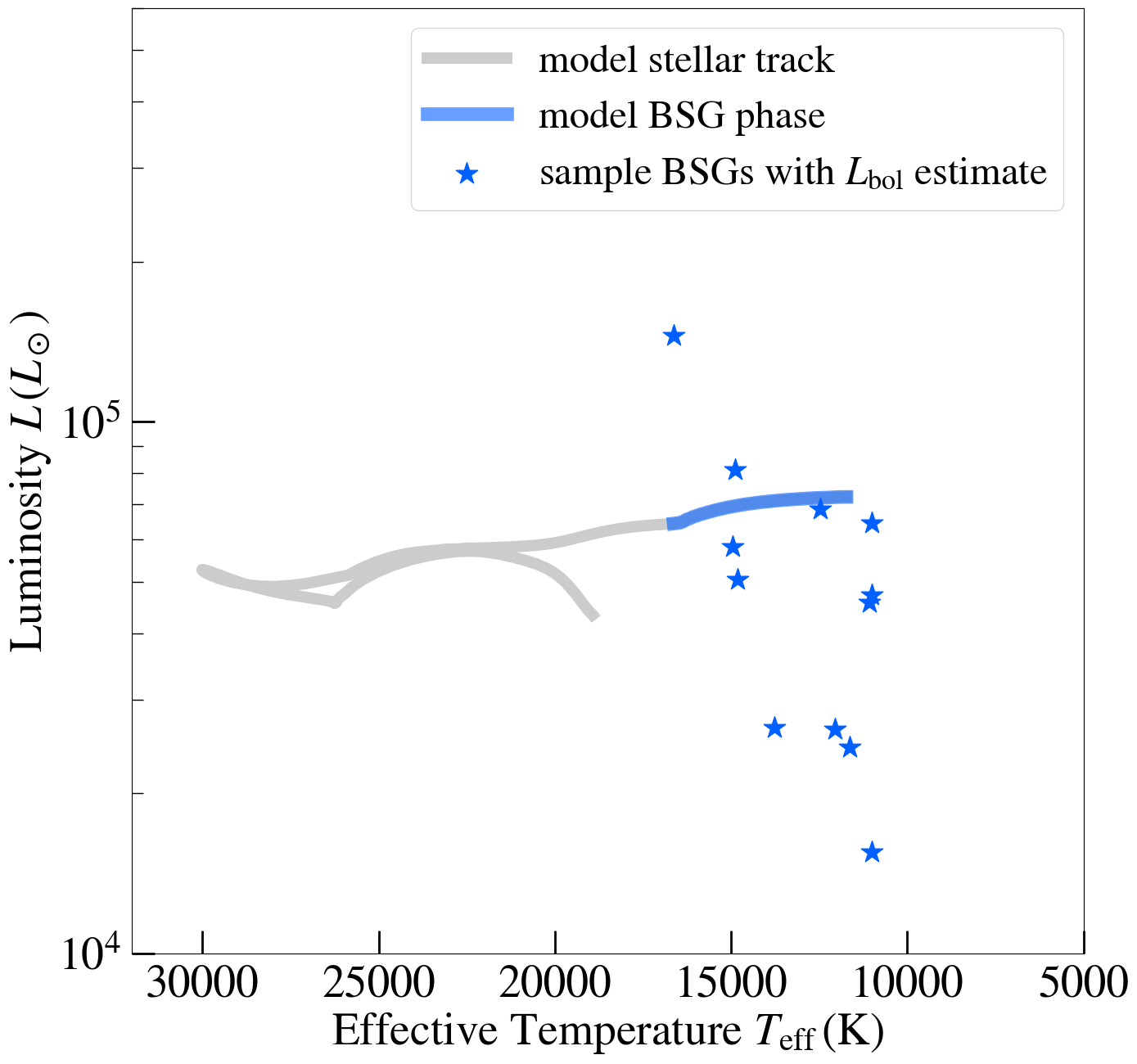}
    \caption{The stellar track of a $15\,M_\odot$ BSG model with a metallicity of $0.008$ on the H--R diagram. This is a central helium burning star formed after a stellar merger \protect\citep{Bellinger2023}, with a BSG lifetime of 1 Myr (highlighted with thick blue line). The positions of 12 BSGs with derived bolometric luminosities in our sample are shown as blue stars.}
    \label{fig:hr_model}
\end{figure}

In this section, we discuss the possible physical mechanisms that may cause the signal we found in our BSG sample and assess their viability in explaining the signal that we discussed in this work.

\subsection{Stellar Winds}

It is believed that stellar winds launched from the surface may cause low-frequency variability in massive stars. These clumpy and inhomogeneous winds are caused by some radiative driving mechanism whose exact details are still under debate \citep{Owocki1984,Puls1996,Puls2006,Puls2008}. They might be responsible for the ``red noise'' signal (see, e.g., \citealt{Aerts2018,Ramiaramanantsoa2018,Krticka2018,Krticka2021,Bailey2024}) or even the large macroturbulence observed in spectroscopy of massive stars (e.g., \citealt{SimonDiaz2017}).

\begin{figure*}
    \centering
    \includegraphics[width=\textwidth]{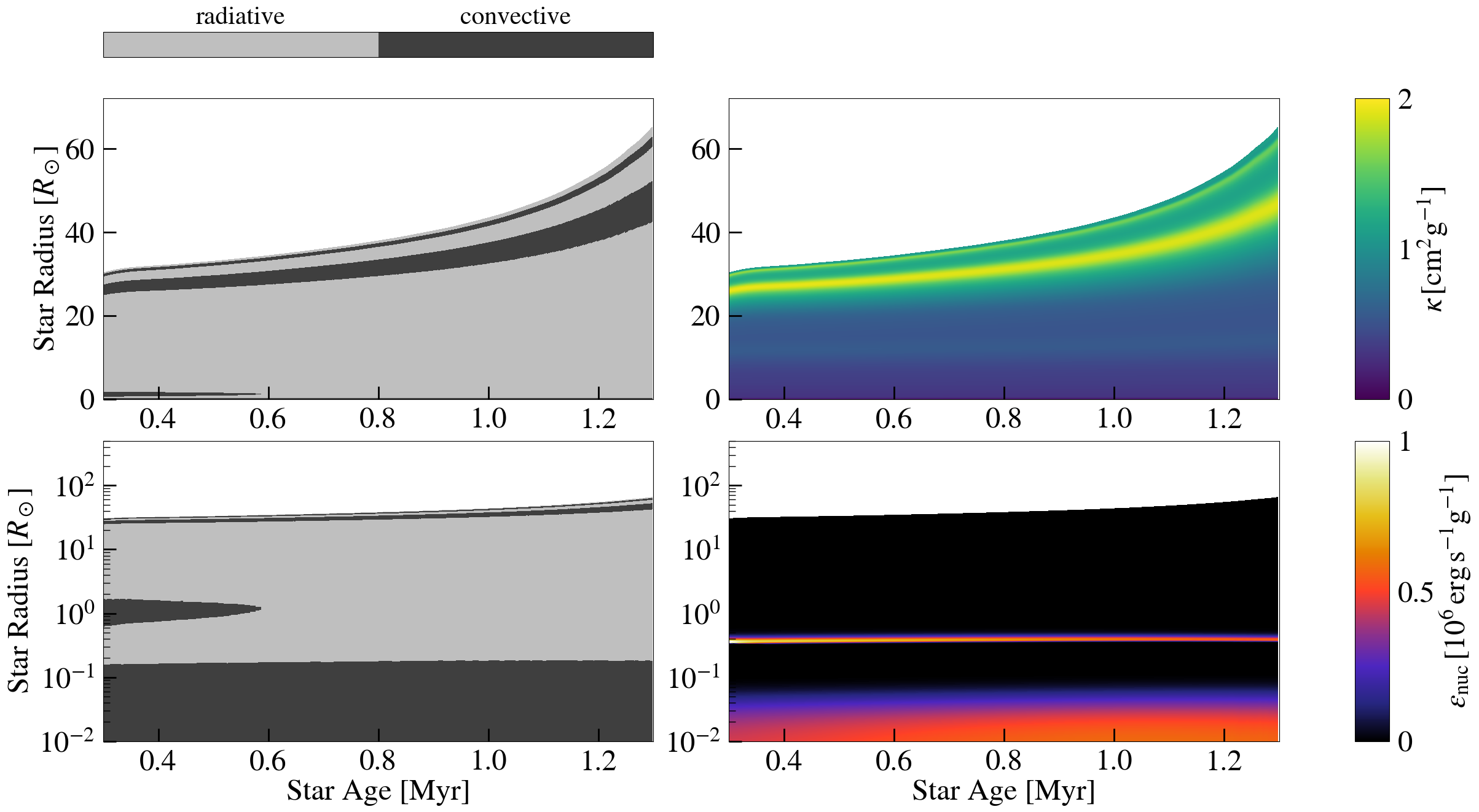}
    \caption{The stellar structure upon radius of our $15\,M_\odot$ model throughout its evolution in the BSG phase (thick blue line in Figure~\ref{fig:hr_model}). The upper panels are in linear coordinates and the lower panels are in logarithm to make the core structures easier to see. {\bf Left}: zone types of the star. We see two subsurface convective zones and a convective core. Before 0.6 Myr there is also a convective zone above the shell-burning region; {\bf Right}: the opacity and nuclear generation rate of the model. The opacity profile shows two major peaks corresponding to the subsurface convective zones, and nuclear energy indicates that the two additional convective regions are caused by shell and core burning.}
    \label{fig:model}
\end{figure*}

\cite{Blomme2011} argue that if stellar winds cause some low-frequency signal, its frequency should be compatible with the stellar rotation frequency at which the winds are launched, and one can then compare the equatorial velocity $v_\mathrm{eq,derive}:=2\pi \nu_\mathrm{peak}R_\mathrm{star}$ derived from the peak frequency of the signal with spectroscopic $v\sin i$ measurements of the star. To test this scenario, we derive the bolometric luminosity $L_\mathrm{bol}$ of 12 BSGs in our sample with reliable {\it G}-band magnitude and extinction measurements from {\it Gaia} \citep{GaiaDR2,GaiaDR3}, with the bolometric correction table provided by the Mesa Isochrones and Stellar Tracks (MIST) grids \citep{Dotter2016,Choi2016}, assuming a distance of 49.97 kpc (the distance of LMC). We then derived their radii $R_\mathrm{bol}$ with the spectroscopic $T_\mathrm{eff}$ from \cite{Serebriakova2023}, and calculated their derived equatorial velocities based on the fitted characteristic frequency in the signal. The results are shown in Table \ref{tab:eq_v}. Despite the uncertainties we introduced in this process, we found that $v_\mathrm{eq,derive}$ for these stars are generally one order of magnitude larger than their measured $v\sin i$, suggesting their inclinations are all less than $\sim10$ degrees. Hence we do not believe the signal is likely caused by stellar winds.

We note, however, whether $v\sin i$ measurements from spectroscopy should be interpreted as stellar rotation is actually unclear. For example, \cite{SimonDiaz2014} and \cite{SimonDiaz2017} point out that macroturbulent velocities may contribute to the total broadening of spectral lines in OB stars at least as equally as rotation, such that the measured $v\sin i$ overestimates the real stellar rotation rate \citep{Schultz2023b}. Nevertheless, this picture only disproves the wind scenario more since it predicts even lower peak frequency compared to what we found from the signal.

\subsection{Subsurface Convection}
\label{sec:conv}

Historically, stochastic and nonperiodic variability is mainly discussed in solar-like oscillators and red giants, in the context of its association with surface convection and granulation \citep{Schwarzschild1975,Michel2008,Chaplin2013,Kallinger2014,Hekker2017}. In massive stars, subsurface convective zones caused by opacity peaks associated with iron and helium ionization may create convective motion in the stellar photosphere, observed as low-frequency photometric variability \citep{Cantiello2009,Cantiello2021}.

To see if these subsurface convective zones are related to our signal, we looked into a $15\,M_\odot$ BSG model
from \cite{Bellinger2023} with a metallicity of 0.008 (a typical value for LMC stars), made with the Modules for Experiments in Stellar
Astrophysics (r23.05.1, \citealt{Paxton2011,Paxton2013,Paxton2015,Paxton2018,Paxton2019,Jermyn2023}). The model starts as a post-merger star with a helium core and a hydrogen envelope, and it evolves to a BSG during its central-helium-burning phase, with a BSG lifetime of 1 Myr (see Figure~\ref{fig:hr_model} for a comparison between the model and our sample on the H--R diagram). While some authors argue that single stellar evolution could also produce BSGs (see, e.g., \citealt{Walborn1987,Weiss1989,Schootemeijer2019,Kaiser2020}), the BSG subsurface structures should be similar in those alternative models, since the occurrence of element opacity peaks is mostly determined by temperature.

In Figure~\ref{fig:model} we show the structure (zone types, opacity and nuclear burning rates) of our model throughout the BSG phase. We see two subsurface convective zones, corresponding to the opacity peaks caused by helium and iron ionization, at $\log(T)=4.7$ and $5.3$. We note, however, that the subsurface convective zones in our model are much deeper than those found by \cite{Cantiello2009}, who mostly concerned less evolved main-sequence stars. Our convective zones are typically $\gtrsim0.5\,R_\odot$ below the stellar photosphere, one order of magnitude deeper than their findings (see Figure~2 in \citealt{Cantiello2009} for comparison). This means subsurface convection is less likely to be observed as photosphere variability for BSGs. Nevertheless, as we used mixing length theory (MLT) to treat convection in our 1D stellar evolution model, we may seriously underestimate the spatial extension of convection, as confirmed in some recent 3D simulations on subsurface convection \citep{Schultz2022,Schultz2023a,Schultz2023b}. In reality, the subsurface convective zones may extend to the surface to be observed.

\begin{figure*}
    \centering
    \includegraphics[width=\textwidth]{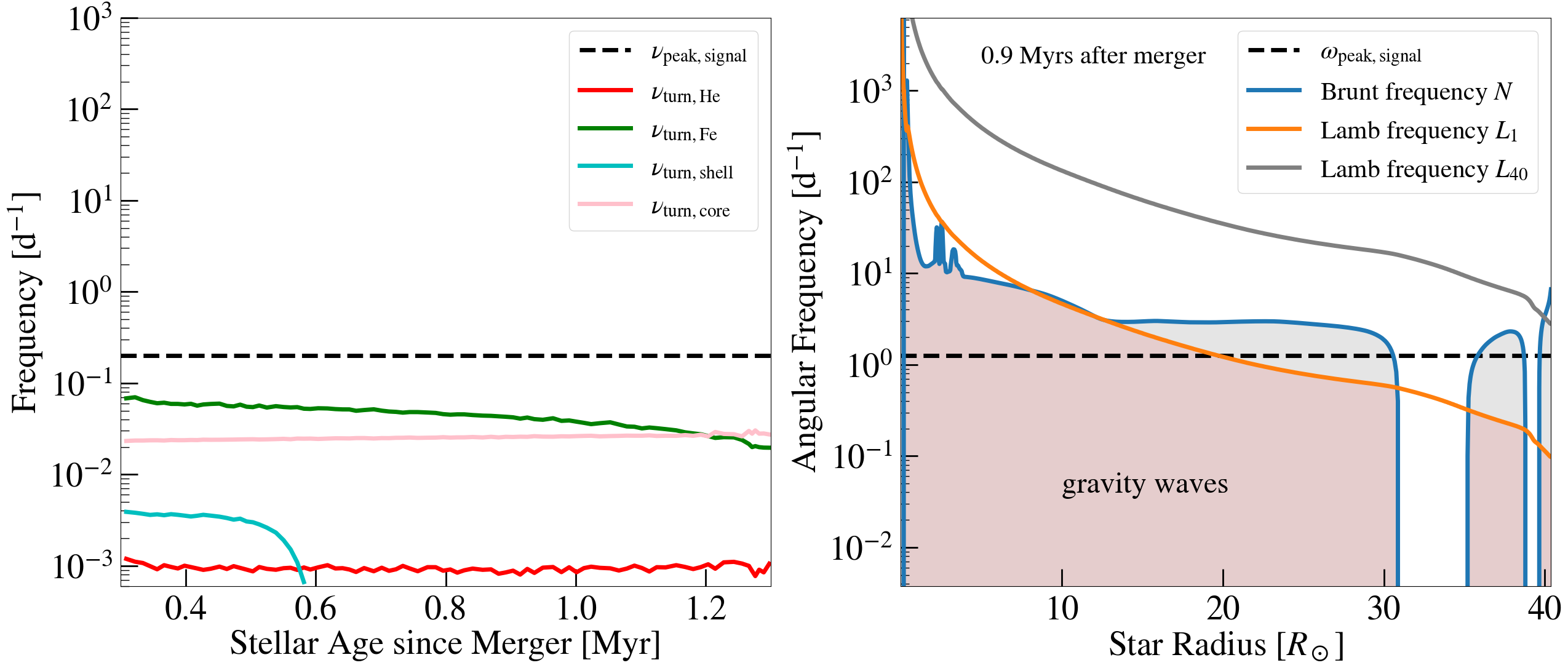}
    \caption{{\bf Left}: The convective turnover frequencies for the convective zones in our BSG model. The frequencies corresponding to the iron opacity peak convective zone and core convection are the closest to the peak frequency of the signal. {\bf Right}: The propagation diagram for the stellar model at $0.9$ Myr after merger, where the colored/gray shaded regions show the frequencies where $l=1$ and $l=40$ gravity waves can propagate. The regions where the Brunt frequency vanishes are the two subsurface convective zones.}
    \label{fig:frequencies}
\end{figure*}

If subsurface convection causes low-frequency variability in some signals, \cite{Cantiello2021} points out that the characteristic frequency of the signal should match the convective turnover frequency on top of the subsurface convection zone. In the left panel of Figure~\ref{fig:frequencies} we plot these frequencies of the convective zones in our BSG models (including two subsurface zones caused by opacity peaks, a convective zone above the burning shell before 0.6 Myr, and burning the convective core, see Figure~\ref{fig:model} for reference), calculated by $\nu_\mathrm{turn}=(2\pi\alpha_\mathrm{MLT}\bar{H}_\mathrm{P}/\bar{v}_\mathrm{c})^{-1}$ (with $\alpha_\mathrm{MLT}=2.0$, see Section 3 in \citealt{Cantiello2021} for details), along with the typical peak frequency in our signal $\nu_\mathrm{peak}=0.2\,\mathrm{d}$. The results seem to favor the convective zone related to iron opacity peak, whose turnover frequency on top is only lower than the signal frequency by a factor of a few, and the inconsistency may only be caused by the uncertainty of $\alpha_\mathrm{MLT}$ \citep{Cantiello2021}. The frequencies of other convective zones are either too low, or the zones themselves are too deep to create any surface convective motion. However, as 3D simulations found that convective zones could extend spatially to larger scales than MLT predictions \citep{Schultz2022}, we should keep in mind that the two subsurface convective zones may actually merge together, and make our MLT calculations unreliable.

In addition, we note that (sub)surface convection and granulation are mostly discussed to explain red noise signals (see, e.g., \citealt{Kallinger2014}), which is different from the characteristic shape of our signal, as the former usually levels off at zero frequency, where our amplitude spectra vanishes. Physically, one would not expect incoherent convective motion to have some longest cutoff periodicity, creating a decay of power toward lower frequency.

The sector-to-sector time variability of our signal is also difficult to be explained by convection. 3D hydrodynamical simulations have predicted that (sub)surface convection usually leads to a steady-state time granulation-like form of SLF variability \citep{Schultz2022}, unlike the long-term modulation timescales on the order of weeks to months in our signal. This means convective motion alone is not satisfactory to explain the shape and variability of our signal in general.

\subsection{Internal Gravity Waves and Pulsating g-modes}
\label{sec:waves}

Internal oscillations have also been suggested as origins of low-frequency photometric variability for massive stars. These oscillations could be waves that propagate to the stellar surface \citep{Cantiello2009,Bowman2019,Ratnasingam2020}, or pulsating modes that are trapped inside the stars \citep{Shiode2013}. They can be driven by the $\kappa$-mechanism associated with iron opacity peaks, which is responsible for nonstable pulsations in $\beta$ Cephei stars and slowly pulsating B stars (see, e.g., \citealt{Pamyatnykh1999,Dupret2001}), or excited stochastically by turbulent convective motion (see, e.g., \citealt{Lecoanet2021,Thompson2023}). In BSGs with convective zones caused by the iron opacity peak, in principle both mechanisms could be related. Nevertheless, the stochasticity in our signal seems to favor convective excitation, and recent 3D simulations showed that modes excited this way can have lifetimes as short as $\sim 10$ days \citep{Thompson2023}, consistent with our findings (see Section \ref{sec:stochaticity}).

It has been proposed that several types of waves may propagate in stars that create low-frequency signals, including Rossby waves that are restored by Coriolis force \citep{vanreeth2016,Saio2018} and gravity waves that are restored by buoyancy \citep{Cantiello2021,Schultz2022}. Among them, we point out that damped gravity waves may naturally explain the shape of our signal at the low-frequency end: with shorter wavelengths at longer periods, gravity waves are preferentially damped at lower-frequency when they propagate in stellar radiative zones, creating smaller photometric amplitudes when they reach the surface. This is exactly what we found from our signal, and has been verified by recent 3D simulations on wave propagation (see Figure 2 in \citealt{Anders2023}). 

\cite{Goldreich1990} showed that if gravity waves are excited by convective motion, they should have a peak power around the convective turnover frequency on top of that convective zone. We hence showed these frequencies calculated for the different convective zones from our BSG model in the left panel of Figure~\ref{fig:frequencies}, compared to the typical peak frequency in our signal $\nu_\mathrm{peak}=0.2\,\mathrm{d}$. We see that the convective turnover frequencies on top of the iron-opacity convective zone and the convective core are both within an order of magnitude lower than the observed peak frequency of the signal (an uncertainty tolerated by the choice of $\alpha_\mathrm{MLT}$, see Section \ref{sec:conv}), suggesting they might generate a spectrum of waves that can be interpreted as our signal. We note, however, \cite{Anders2023} point out that waves generated from the core cannot get to the observed amplitude when they reach the surface, though in their simulations, they did not have other convective zones above the core, which may amplify or further damp the wave amplitude and it travels through them.

To see whether gravity waves excited by convection can propagate, we show a propagation diagram (i.e., the parameter space of the angular frequencies and locations inside a star where waves can propagate) in the right panel of Figure~\ref{fig:frequencies}, for our BSG model at a typical stellar age of $0.9\,\mathrm{Myr}$. Gravity waves can only propagate in regions where their frequencies are below both the Brunt-V\"ais\"al\"a frequency \citep{Vaisala1925} and the Lamb frequency \citep{Unno1989}. We see that for $l=1$ waves, the cutoff frequency above the iron opacity peak is below the peak frequency we find in our signal, which means waves are at our peak frequency evanescent. Nevertheless, \cite{Cantiello2009b} and \cite{Shiode2013} demonstrated that gravity waves excited by subsurface convection can have very high degrees ($l\gtrsim 30$), whose cutoff frequencies are above the peak frequency of our signal (see Figure~\ref{fig:frequencies} for the propagation of an example $l=40$ wave), such that they may propagate to the surface. However, these high-degree oscillations may only produce small amplitudes because of geometric cancellation effects (see, e.g., \citealt{Aerts2010book}), and it is questionable whether they can produce the observed amplitude of our signal (up to a few percent of stellar luminosity).

The sector-to-sector time variability of our signal is compatible with an origin of stochastically excited waves, as the long-period beating patterns among many superimposed waves of various spatial scales would lead to modulation timescales of months to years. An additional constraint from the signal is that there are no individual modes identified for any of our BSGs around the characteristic peak frequency. \cite{Bellinger2023} proposed that if BSGs are formed by mergers, the $l=1$ g-mode period spacing would be $\Delta\Pi_1\approx 20-30\,\mathrm{min}$. To identify individual g modes around $\nu_\mathrm{peak}\sim0.2\,\mathrm{d}^{-1}$, this requires the frequency resolution to be less than
$\Delta\nu=\nu_\mathrm{peak}^2\Delta\Pi\approx10^{-3}\,\mathrm{d}^{-1}$, marginally below the resolution of the full TESS light curves of our sample. For higher-degree g modes, their geometric cancellation effects make it impossible to identify them from photometry. Therefore, the lack of identified modes from the signal still could not exclude the possibility that they are a spectrum of individual g-modes. Nevertheless, we expect further observations with increased photometric precision (such as those with the PLATO mission; \citealt{Miglio2017}) and observations seeking for spectroscopic time variability will help to distinguish these modes.

We note, however, that our wave analysis is based on timescale arguments on this single 1D stellar evolution model. More detailed investigation including solving oscillation modes/3D modeling of convection should be carried out to verify this explanation in the future.

\label{sec:modes}

\section{Conclusion}
\label{sec:conclusions}

In this manuscript, we analyzed TESS light curves for 20 blue supergiants (BSGs) in the LMC, and found a characteristic signal in the low-frequency ($f\lesssim 2\,\mathrm{d}^{-1}$) range for all our targets. The signals show strong stochasticity across different TESS sectors, yet their full amplitude spectrum show a peak frequency around $0.2\,\mathrm{d}^{-1}$, below which the amplitude tends to zero. We were able to fit the signal with a modified Lorentzian profile (Equation \ref{eq:fit}), and the fitting parameters have no strong correlations with spectroscopic parameters measured for these systems.

We compared our signals to those obtained from a limited sample of hotter OB stars, yellow supergiants, alpha Cygni variables and red supergiants. We see similarities between our signal and the low-frequency variability of hot OB stars, yellow supergiants and some alpha Cygni variables. This comparison, while limited, may suggest the origins of these signals are similar.

We discussed three possible mechanisms that may explain this signal: stellar winds launched by rotation, subsurface convective motion, and internal oscillations of the star. The spectroscopically measured $v\sin i$ seems to disfavor the wind mechanism, as it would produce too low a characteristic frequency compared to the signal. We created a BSG model to test the latter two scenarios, and we found that the turnover frequency on top of the convective zone caused by the iron opacity peak is consistent with the peak frequency of the signal, despite the uncertainties introduced by the choice of mixing length parameter. While convective motion in this zone may be interpreted as the signal, it might be too deep to be directly observed in the photosphere, and such motions typically do not predict the shape and short-term time variability of our amplitude spectra. High-order, damped gravity waves excited from this convective zone seem to be a better explanation, which explains the peak frequency, the shape of the signal and the short-term time variability, though this picture will require more thorough investigation.

We restricted our sample to 20 BSGs with reliable TESS time series and spectroscopic parameters in the LMC. While this gives a clean sample to draw away the signal, we do not know whether it is still present for higher-metallicity BSGs. In future work, we will extend our analysis to more Galactic BSGs. A greater population of samples may also help us to seek potential correlations between the signal and stellar parameters.

Our theoretical analysis is somewhat limited by the single BSG model we chose, and future work should also focus on more detailed modeling of the BSGs in our sample.

\begin{acknowledgments}
We thank the anonymous referee for the constructive report that helps to improve this work. We thank Dominic Bowman, Jim Fuller, Matteo Cantiello, Stephen Justham, Lars Bildsten, Yanqin Wu and Norbert Langer for useful discussions. This work was initiated at the Kavli Summer Program in Astrophysics 2023, hosted at the Max Planck Institute for Astrophysics. We thank the Kavli Foundation and the MPA for their support. CJ gratefully acknowledges support from the Netherlands Research School of Astronomy (NOVA).
\end{acknowledgments}

\vspace{5mm}
\facilities{ Some of the data presented in this paper were obtained from the Mikulski Archive for Space Telescopes (MAST) at the Space Telescope Science Institute. The specific observations analyzed can be accessed via \dataset[https://doi.org/10.17909/df38-ax53]{https://doi.org/10.17909/df38-ax53}. STScI is operated by the Association of Universities for Research in Astronomy, Inc., under NASA contract NAS5–26555. Support to MAST for these data is provided by the NASA Office of Space Science via grant NAG5–7584 and by other grants and contracts.
This work presents results from the European Space Agency (ESA) space mission Gaia. Gaia data are being processed by the Gaia Data Processing and Analysis Consortium (DPAC). Funding for the DPAC is provided by national institutions, in particular the institutions participating in the Gaia MultiLateral Agreement (MLA). The Gaia mission website is https://www.cosmos.esa.int/gaia. The Gaia archive website is https://archives.esac.esa.int/gaia.}

\software{ This research made use of Astropy,\footnote{http://www.astropy.org} a community-developed core Python package for Astronomy \citep{astropy:2013, astropy:2018, astropy:2022}. This research made use of Lightkurve, a Python package for Kepler and TESS data analysis \citep{Lightkurve}. This work has also made use of the LATTE \citep{LATTE} matplotlib \citep{Hunter2007}, numpy \citep{Harris2020}, and SciPy \citep{scipy} packages and the MESA \citep{Paxton2011,Paxton2013,Paxton2015,Paxton2018,Paxton2019,Jermyn2023} software and SDK (version 22.6.1: https://doi.org/10.5281/zenodo.7457723).}

\appendix

\section{Impact of light curve reduction on low frequency signal}
\label{app:spoc}

\begin{figure}
    \centering
    \includegraphics[width=0.9\columnwidth]{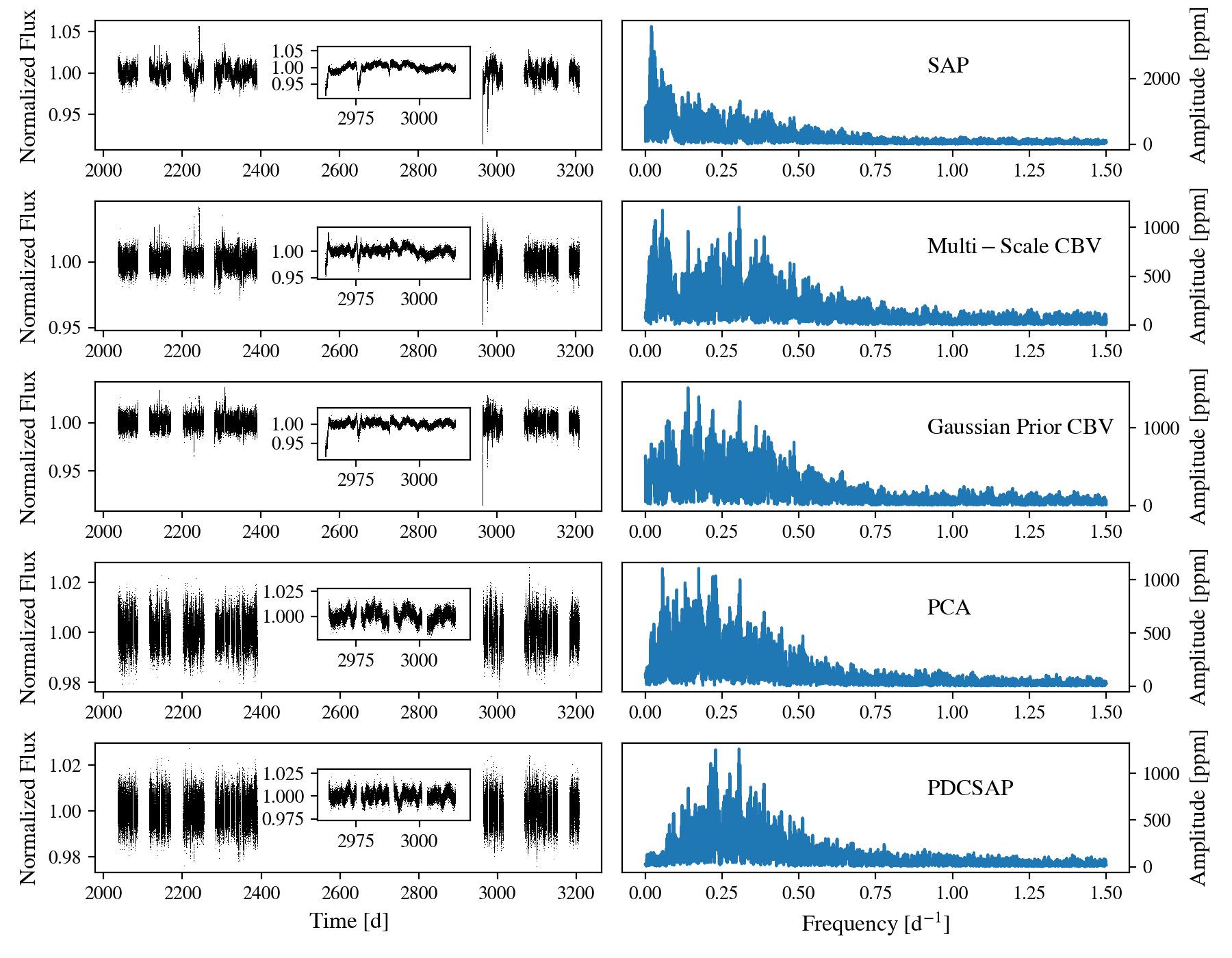}
    \caption{Example light curves (left) and accompanying amplitude spectra (right) for TIC~425084139, with different reduction methods.}
    \label{fig:detrended_periodograms}
\end{figure}

In this work, we made use of the {\it TESS} two minute cadence pre-search data conditioning simple aperture photometry, or PDCSAP, data downloaded from MAST that was reduced and treated using the SPOC pipeline \citep{Jenkins2016}. This pipeline removes systematic trends introduced by various instrumental effects. Specifically, the SPOC pipeline does bias, flat-fielding and background corrections and removes trends that are introduced by Earth-shine events, thermal brightening events, loss-of-fine-pointing events, changes in pixel sensitivity, and regular momentum dumps. Generally, these events introduce noise at low frequencies that can hide astrophysical signals at low frequencies. To mitigate these events, the SPOC pipeline utilizes co-trended basis vectors (CBVs) which encapsulate the common trends per CCD channel that are present in a sample of strategically selected photometrically quiet stars \citep{Kinemuchi2012,Stumpe2012}. These CBVs are fit to each star individually, and the pipeline decides which removal method is best. To investigate the impact that the removal of these trends has on the resulting signals in the periodogram, we investigate five light curves with different reductions. In particular, we investigate the raw SAP flux data, SAP flux data that has been corrected using co-trending basis vectors two different application methods (e.g. multi-scale regularisation and single-scale Gaussian prior), a custom pixel-level principal component analysis (PCA) de-trending, and the SPOC produced PDCSAP flux data. In the cases that use CBVs, we only use the first five CBVs. The different light curves and their accompanying periodograms are shown in the left and right columns of Fig.~\ref{fig:detrended_periodograms}, respectively. The raw SAP and both CBV corrected light curves display clear trends that indicate that not all instrumental effects were properly treated. The impact of the remaining trends is seen as a low frequency noise excess below $\sim$0.1 d$^{-1}$ in the top three rows. The PCA corrected light curve shows a clear decrease in the overall noise at the lowest frequencies, however, there are still trends at the beginning and end of observing sectors that are not totally removed by the PCA method (demonstrated in the insets), resulting in some remaining low frequency noise. We note that only the PDCSAP SPOC light curve with the full calibrated reduction process sufficiently removes instrumental trends to explore low frequency variability.

\section{TESS Light Curves and periodogram for full sample}
\label{app:all_data}

In Figures \ref{fig:31179797} to \ref{fig:179038240} we present the light curves and periodograms of all our BSG sample. They show a universal pattern in the low-frequency range.

\figdata{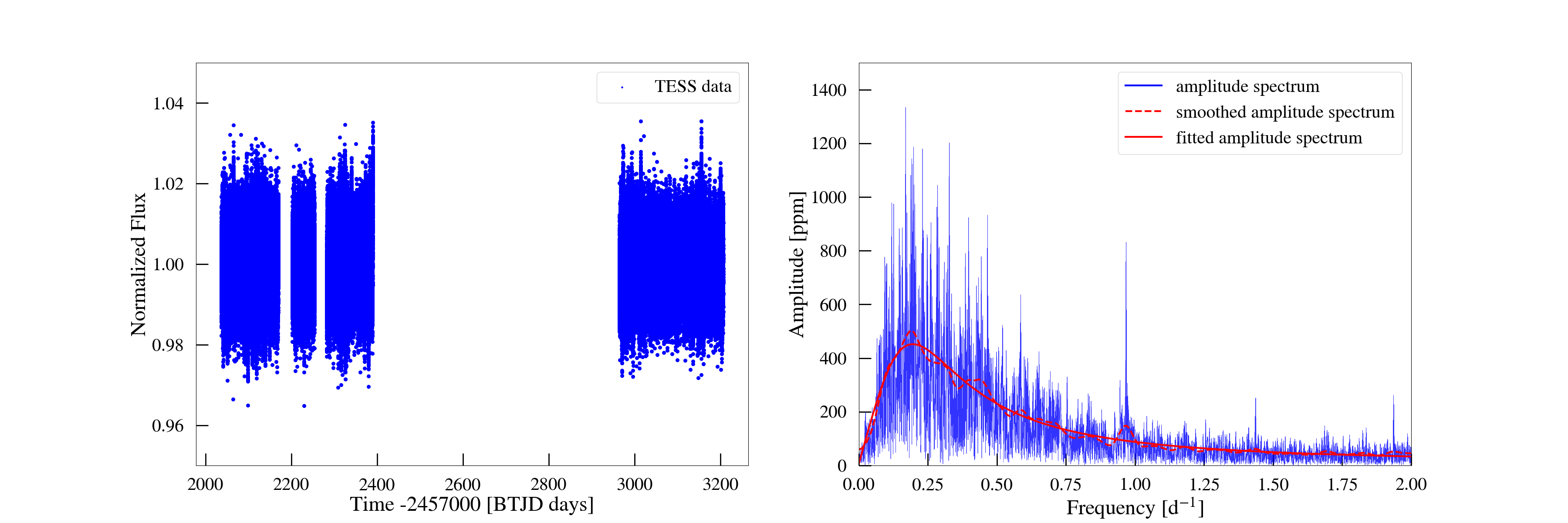}{SK -68 53}
\figdata{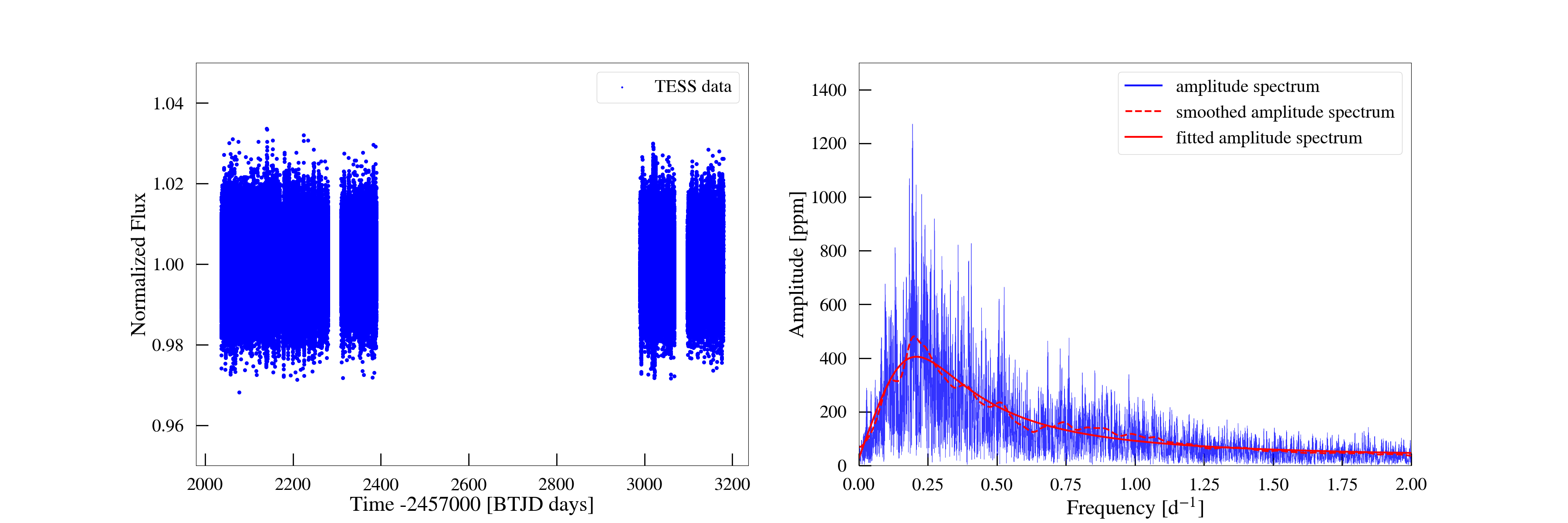}{SK -66 125}
\figdata{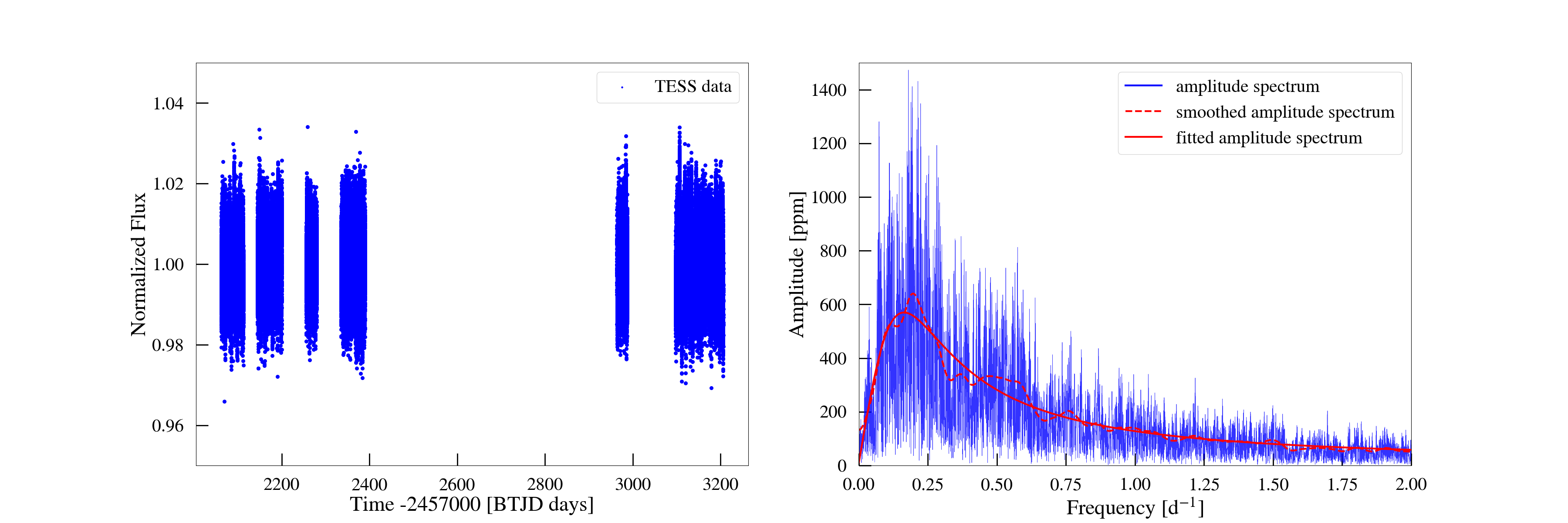}{SK -67 283}
\figdata{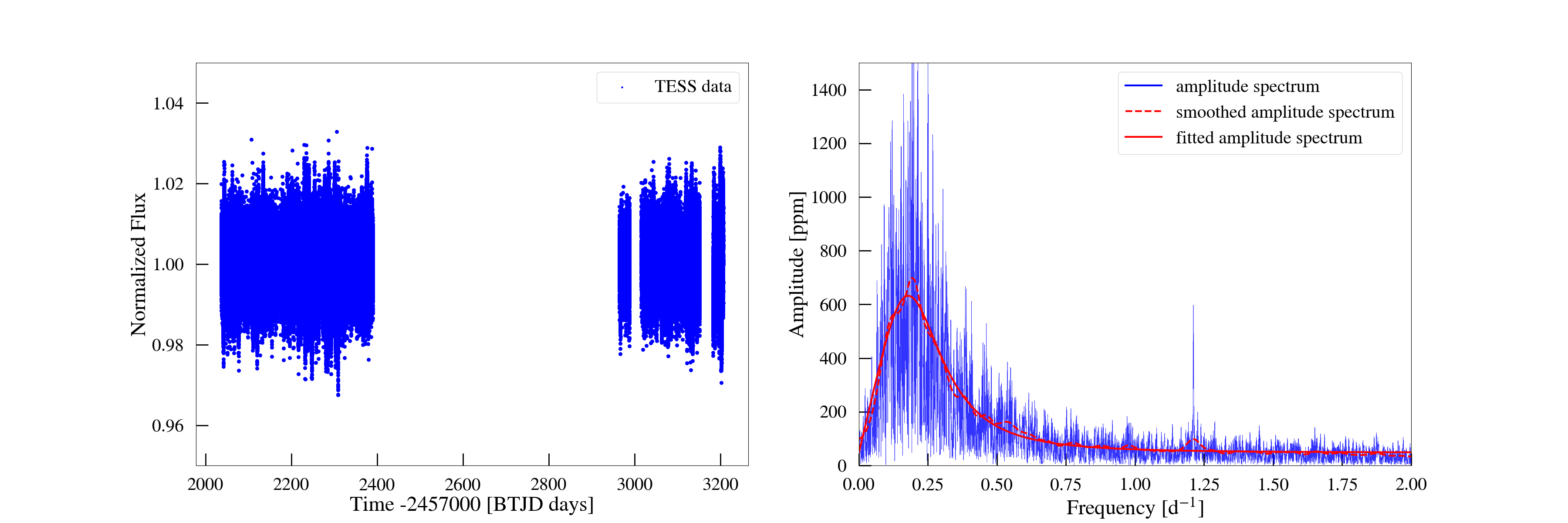}{SK -69 31}
\figdata{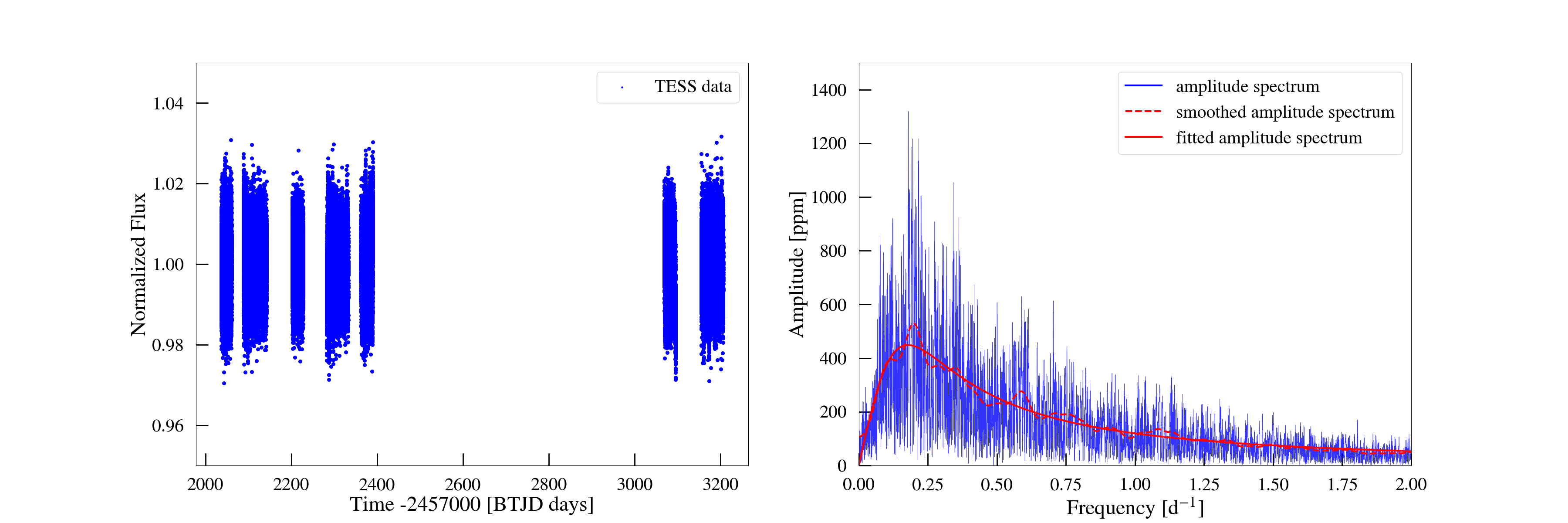}{SK -67 275}
\figdata{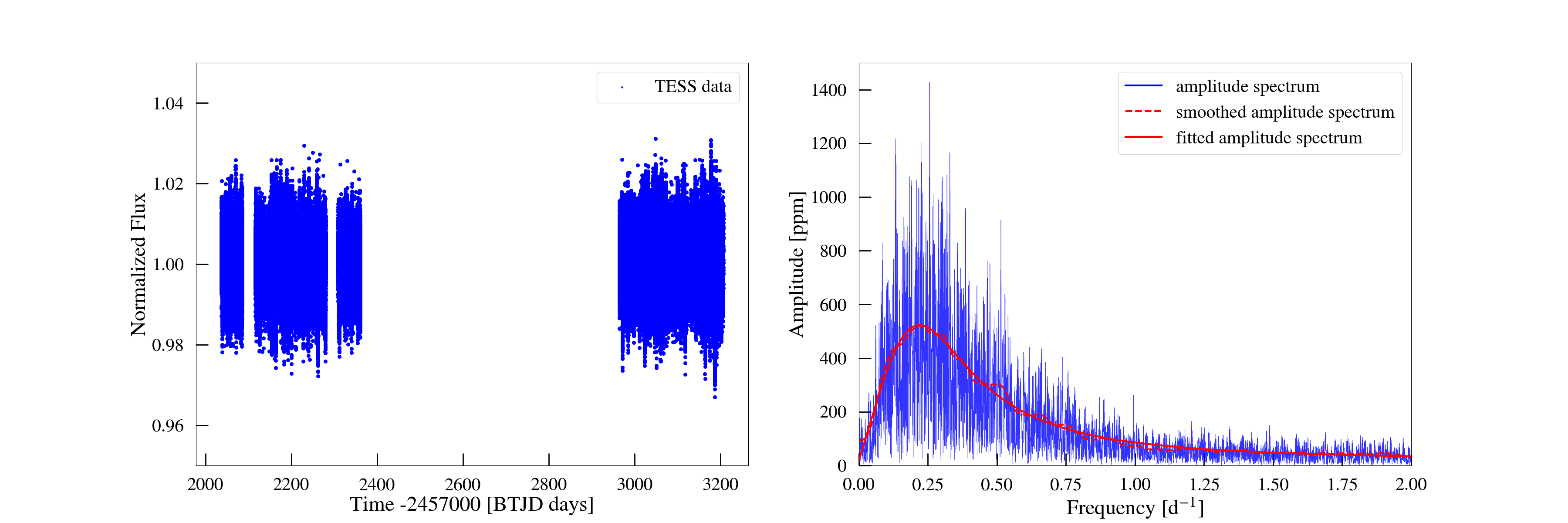}{SK -66 142}
\figdata{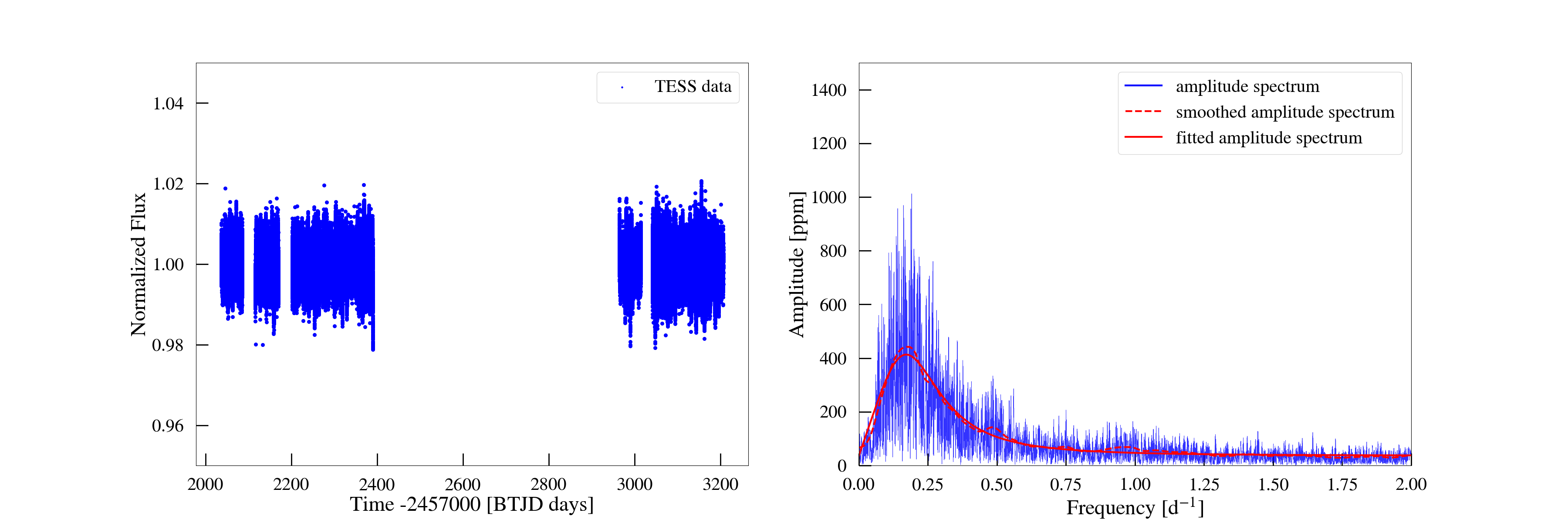}{HD 269639}
\figdata{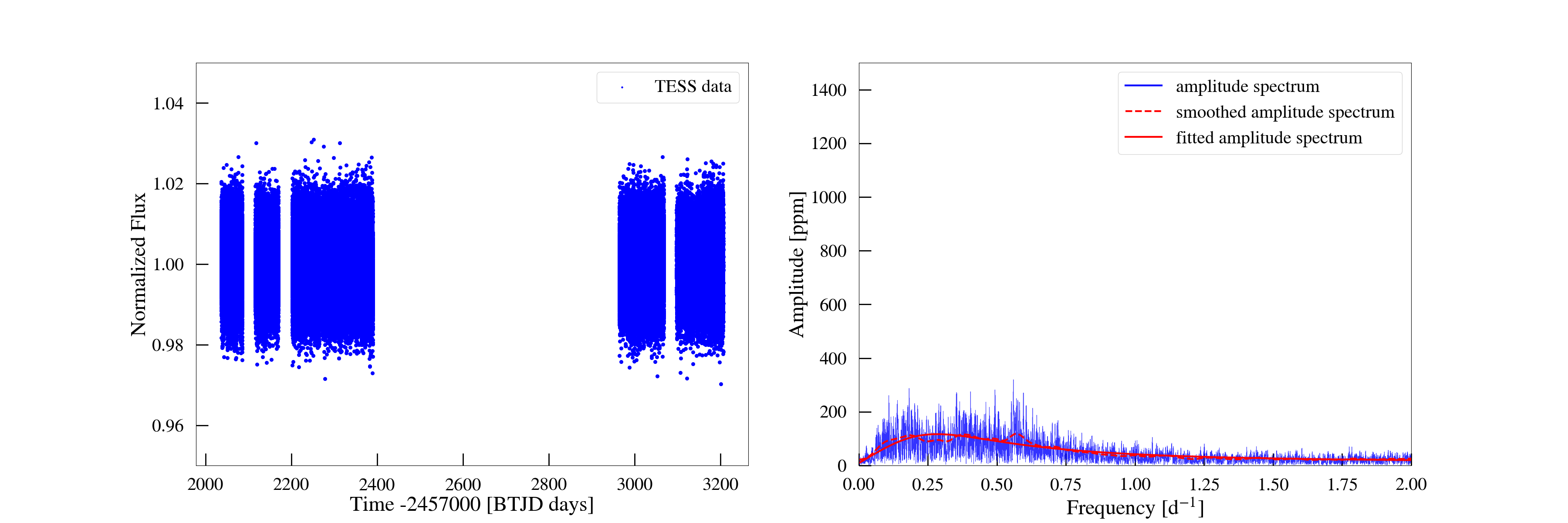}{SK -67 7}
\figdata{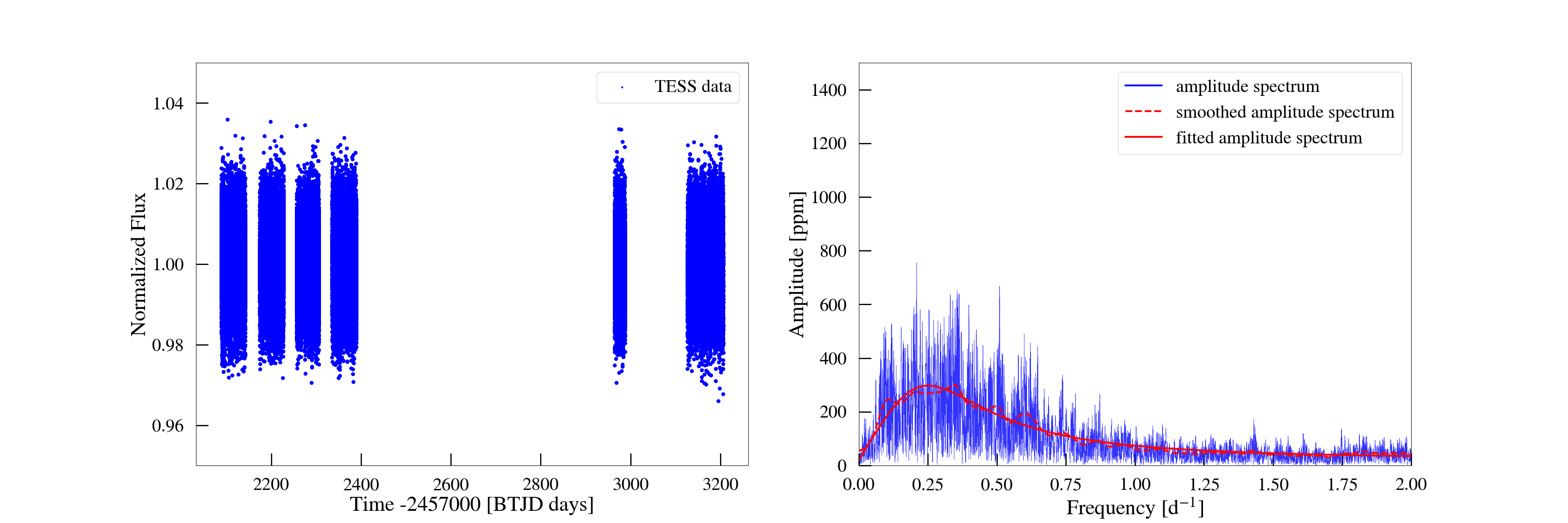}{SK -67 279}
\figdata{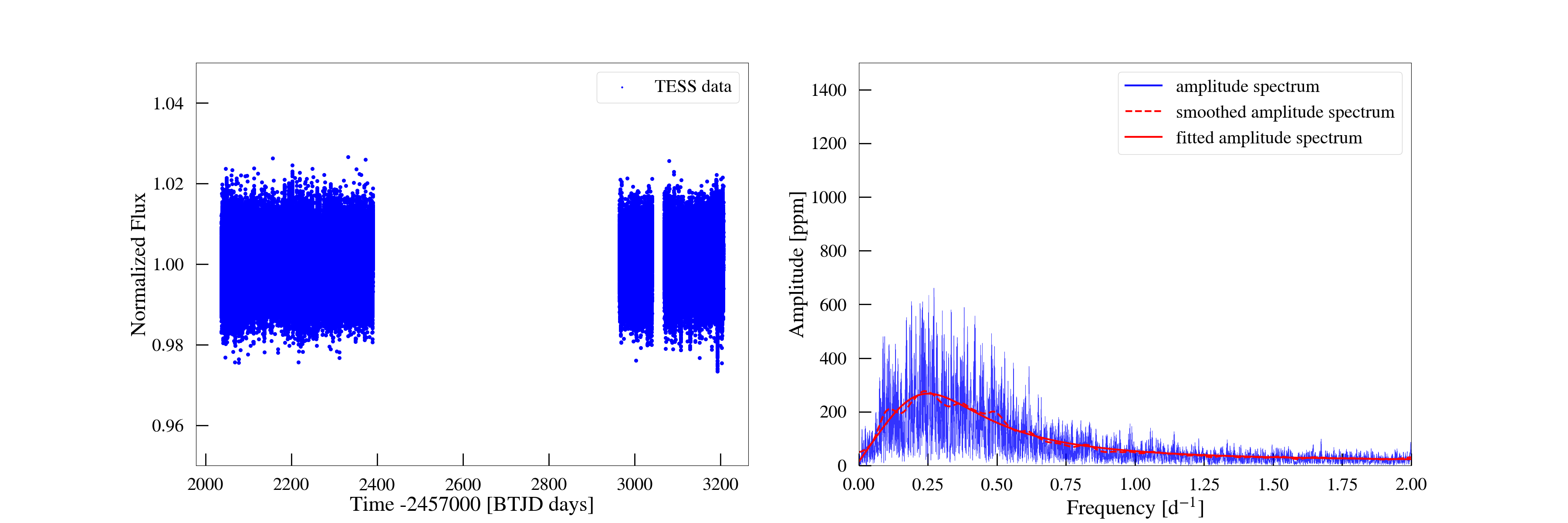}{SK -70 31}
\figdata{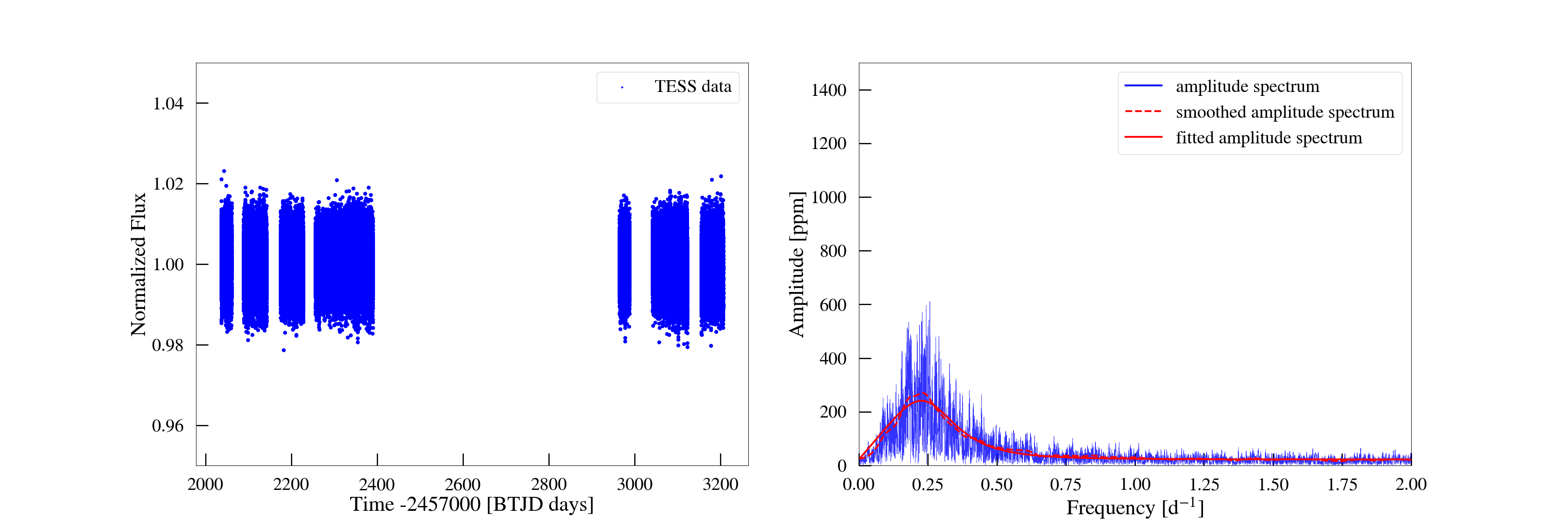}{SK -68 152}
\figdata{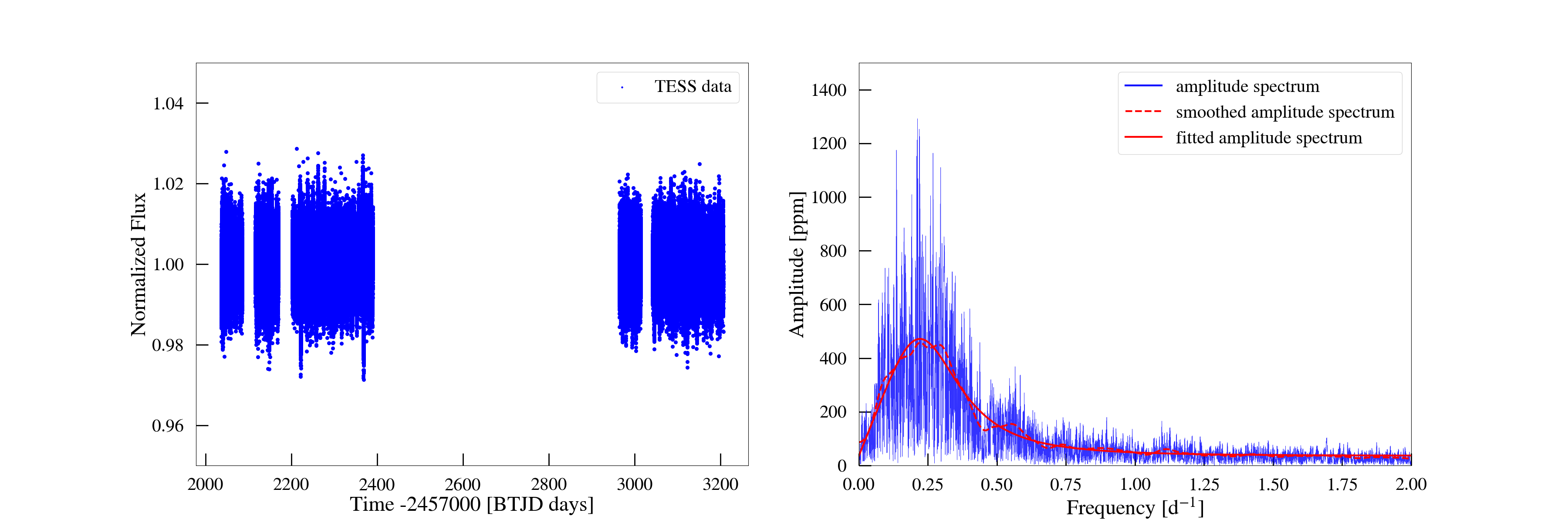}{SK -67 88}
\figdata{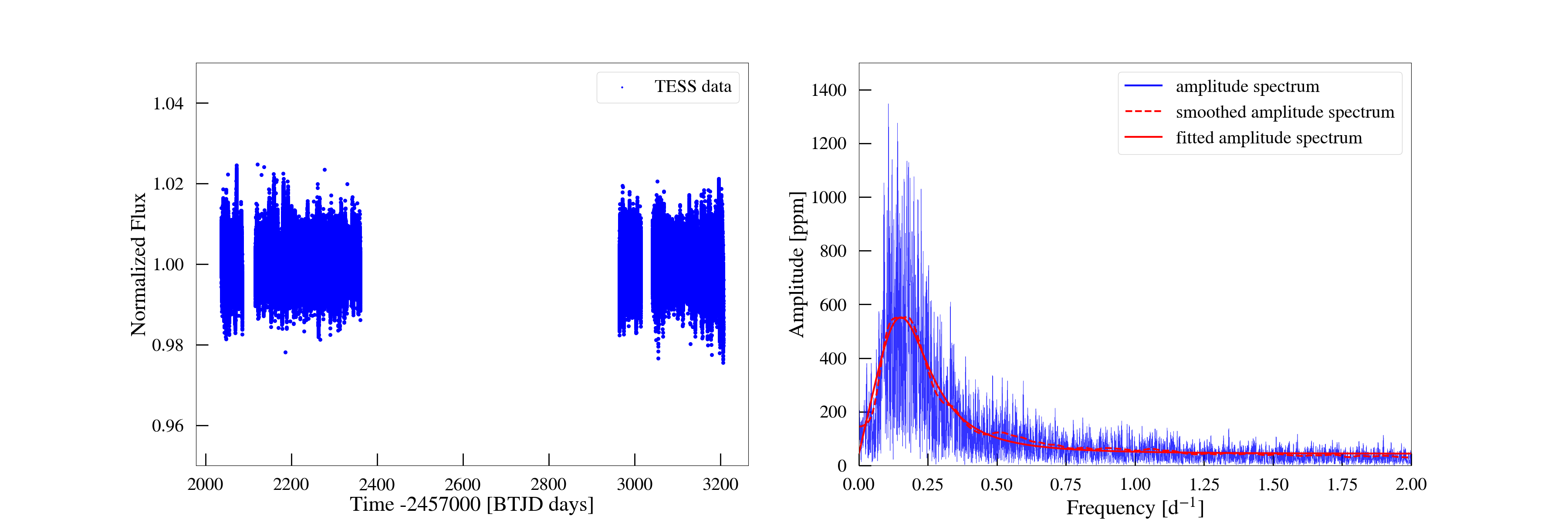}{HD 269721}
\figdata{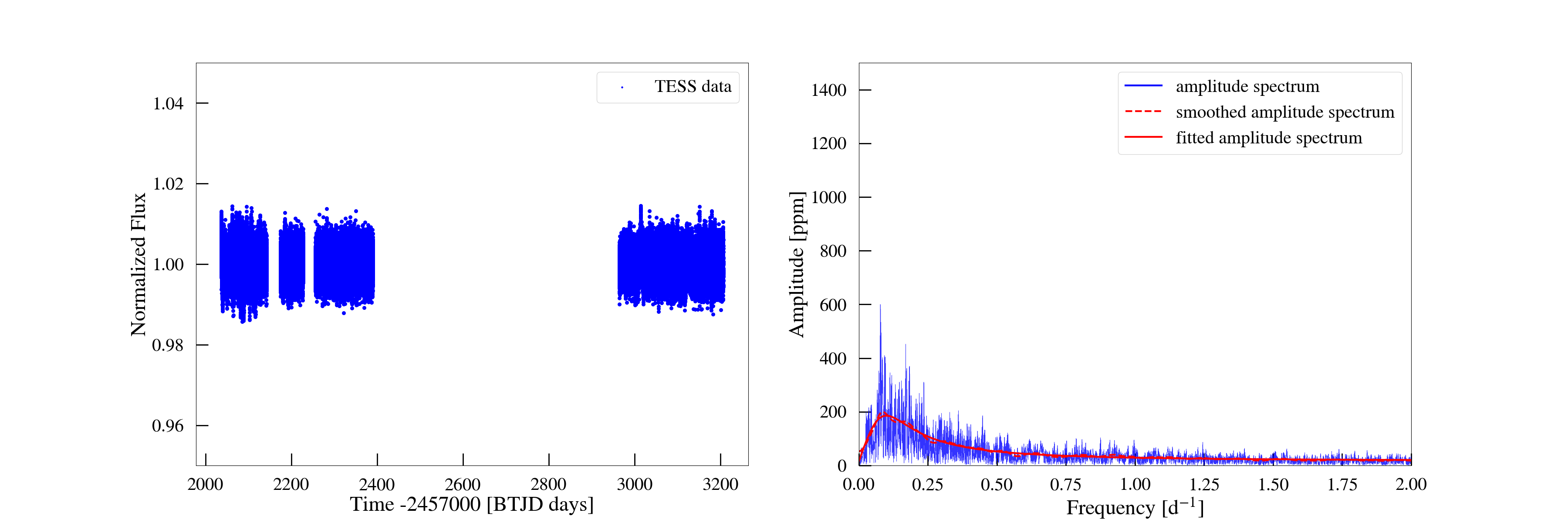}{HD 269510}
\figdata{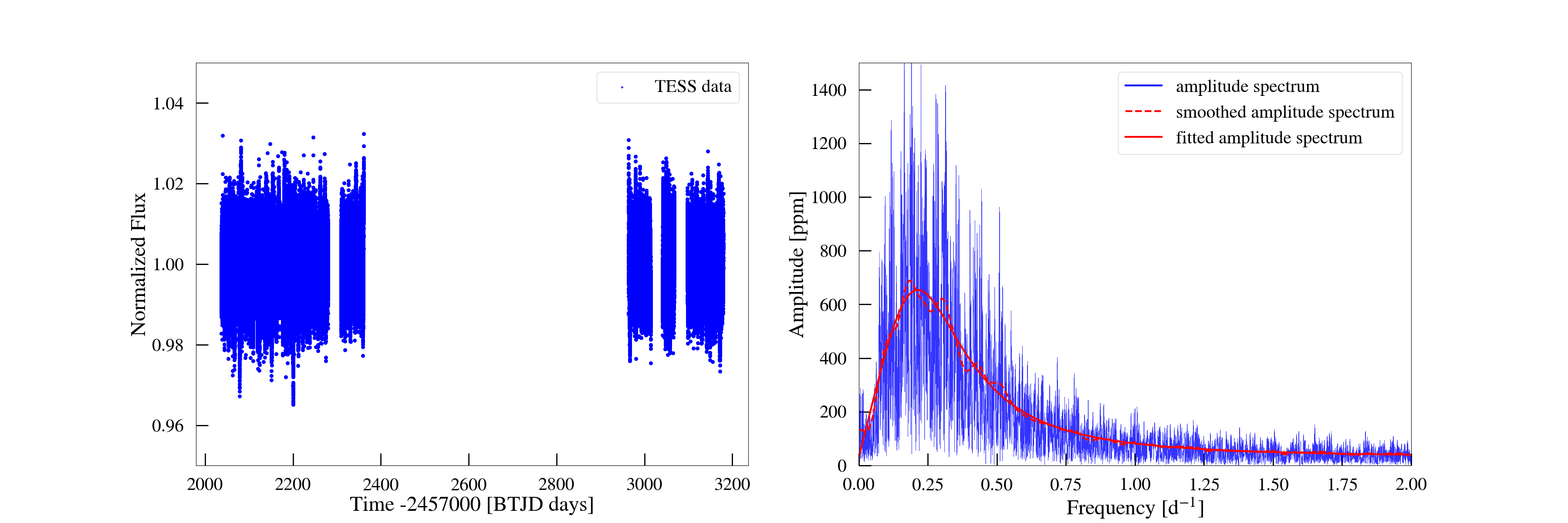}{SK -66 92}
\figdata{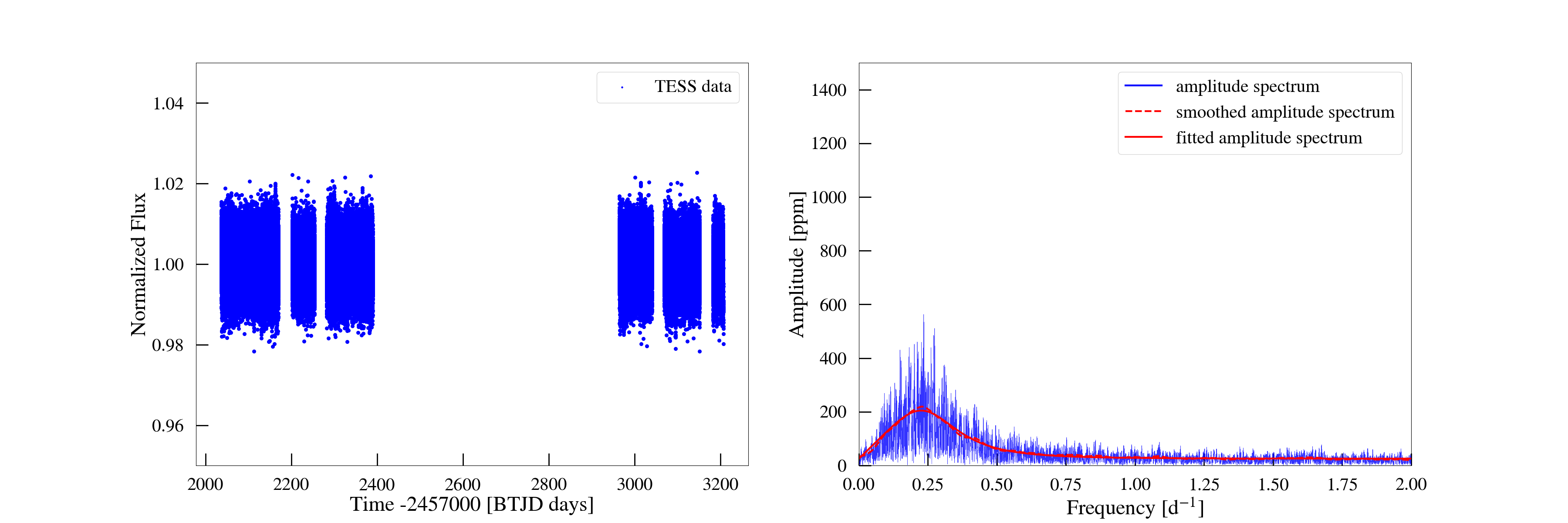}{SK -67 151}
\figdata{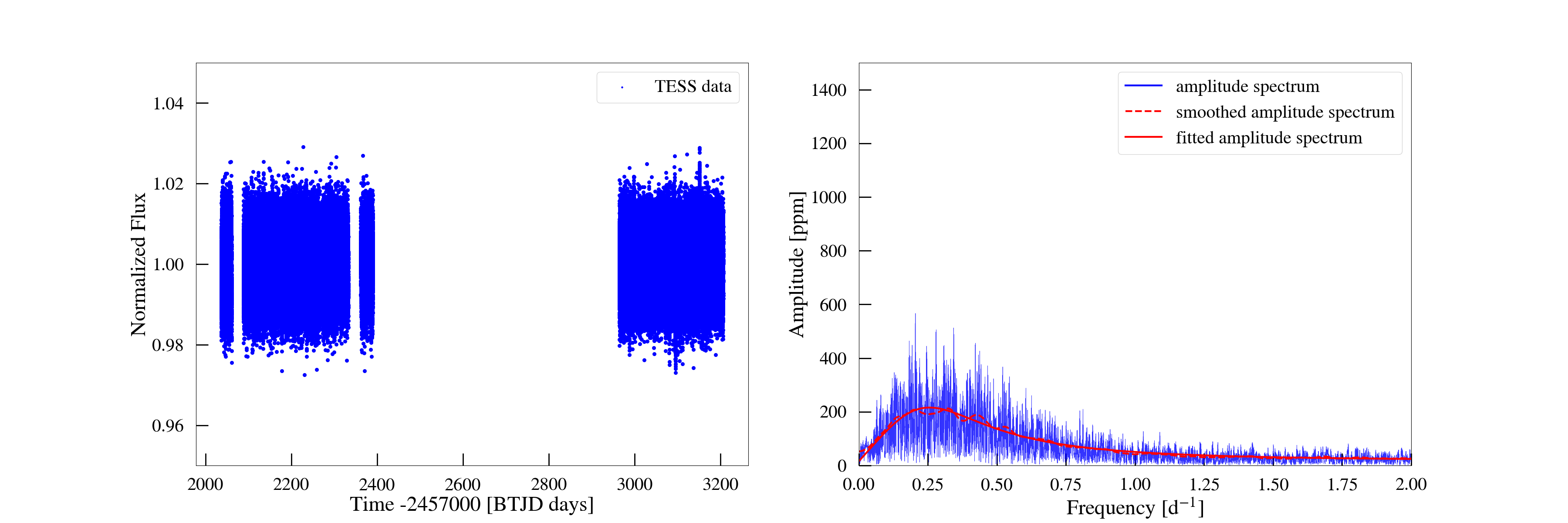}{SK -70 26}
\figdata{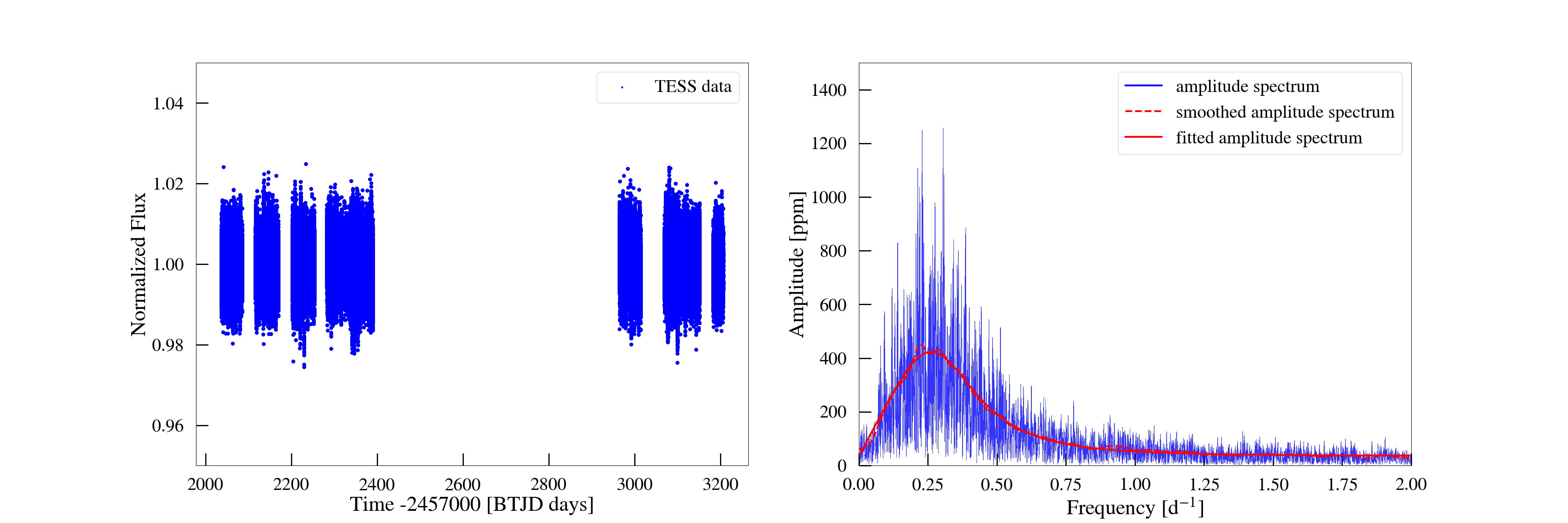}{SK -67 171}
\figdata{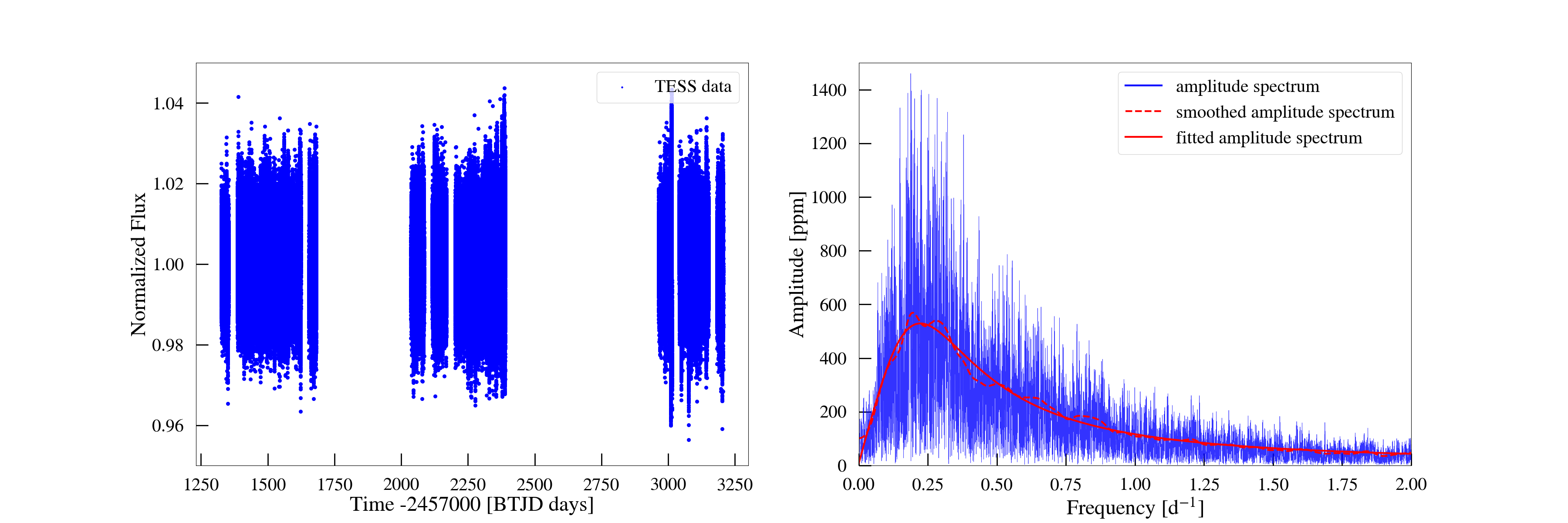}{SK -67 133}
\figdata{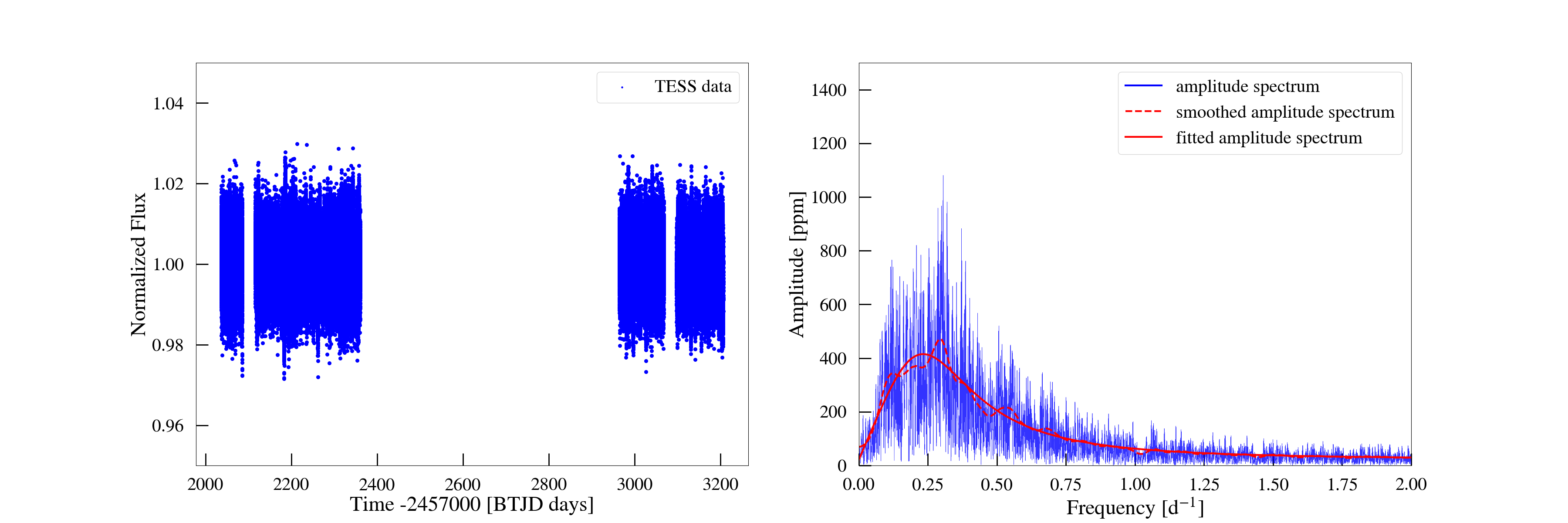}{SK -67 72}

\FloatBarrier
\bibliographystyle{aasjournal}
\DeclareRobustCommand{\VAN}[3]{#3}
\bibliography{BSG}

\end{CJK*}
\end{document}